\documentclass[11pt]{article}

\usepackage{color}
\usepackage{amssymb}
\usepackage{amsmath}
\usepackage[bottom]{footmisc}
\usepackage{mathtools}
\usepackage{acronym}
\usepackage{nicefrac}
\usepackage{graphicx}
\graphicspath{{figs}}
\usepackage{bm}
\usepackage[dvipsnames]{xcolor}
\usepackage[colorlinks=true]{hyperref}
\hypersetup{
	linkcolor = {blue},
	anchorcolor = blue,
	citecolor = blue,
	filecolor = blue,
	urlcolor = blue
}
\usepackage[numbers,sort&compress]{natbib}

\usepackage{authblk}

\usepackage[font=small]{caption}
\usepackage{enumitem}

\usepackage{comment}

\usepackage{xr}
\makeatletter

\newcommand*{\addFileDependency}[1]{
\typeout{(#1)}
\@addtofilelist{#1}
\IfFileExists{#1}{}{\typeout{No file #1.}}
}\makeatother

\newcommand*{\myexternaldocument}[1]{%
\externaldocument[S-]{build/#1}%
\addFileDependency{#1.tex}%
\addFileDependency{build/#1.aux}%
}

\myexternaldocument{supplementary}

    \usepackage{setspace}
    \usepackage[centering,hmargin=2cm,vmargin=2cm]{geometry}
    
    \let\oldabstract\abstract
    \let\oldendabstract\endabstract
    \makeatletter
    \renewenvironment{abstract}
    {%
                   {\list{}{\addtolength{\leftmargin}{1em} 
                            \listparindent 1.5em%
                            \itemindent    \listparindent%
                            \rightmargin   \leftmargin%
                            \parsep        \z@ \@plus\p@}%
                    \item\relax}%
                   {\endlist}%
    \oldabstract}
    {\oldendabstract}
    \makeatother

\begin{document}


\title{Unifying Pairwise Interactions in Complex Dynamics}

\author[1,2]{Oliver M. Cliff}
\author[1,2]{Annie G. Bryant}
\author[2,3]{Joseph T. Lizier}
\author[4,5,6]{Naotsugu Tsuchiya}
\author[1,2]{Ben D. Fulcher}
\affil[1]{School of Physics, The University of Sydney, Camperdown NSW 2006, Australia.}
\affil[2]{Centre for Complex Systems, The University of Sydney, Camperdown NSW 2006, Australia.}
\affil[3]{School of Computer Science, The University of Sydney, Camperdown NSW 2006, Australia.}
\affil[4]{Turner Institute for Brain and Mental Health \& School of Psychological Sciences, Faculty of Medicine, Nursing, and Health Sciences, Monash University, Melbourne, Victoria, Australia.}
\affil[5]{Center for Information and Neural Networks (CiNet), National Institute of Information and Communications Technology (NICT), Suita-shi, Osaka 565-0871, Japan.}
\affil[6]{Advanced Telecommunications Research Computational Neuroscience Laboratories, 2-2-2 Hikaridai, Seika-cho, Soraku-gun, Kyoto 619-0288, Japan.}

\date{\vspace{-5ex}}

\maketitle

\begin{abstract}
    \small
    Scientists have developed hundreds of techniques to measure the interactions between pairs of processes in complex systems.
    But these computational methods---from correlation coefficients to causal inference---rely on distinct quantitative theories that remain largely disconnected.
    Here we introduce a library of 237 statistics of pairwise interactions and assess their behavior on 1053 multivariate time series from a wide range of real-world and model-generated systems.
    Our analysis highlights new commonalities between different mathematical formulations, providing a unified picture of a rich interdisciplinary literature.
    Using three real-world case studies, we then show that simultaneously leveraging diverse methods from across science can uncover those most suitable for addressing a given problem, yielding interpretable understanding of the conceptual formulations of pairwise dependence that drive successful performance.
    Our framework is provided in extendable open software, enabling comprehensive data-driven analysis by integrating decades of methodological advances.
\end{abstract}







A fundamental question in science is how complex dynamics can be characterized by measuring the interactions within a distributed system.
To address this question, many approaches have been developed to measure different types of pairwise interactions from dynamical data.
For example, in neuroimaging, functional connections between pairs of brain regions are quantified through statistical correlations, which mark changes in human behaviors~\cite{bassett2017network} and differ in neurological diseases~\cite{buckner2005molecular}.
In Earth systems science, pairwise causal models have been used to infer mechanistic drivers of natural processes, from the influence of sea-surface temperature on sardine and anchovy populations~\cite{sugihara_detecting_2012} to the atmospheric drivers of air circulation~\cite{runge_inferring_2019}.
And economic analysts have studied the cointegration of paired non-stationary time series, such as stock-market indices and their associated future contracts, to infer a significant coupling for building econometric models~\cite{engle1987co}.

\begin{figure}[t!]
    \centering
     \includegraphics[width=.92\textwidth]{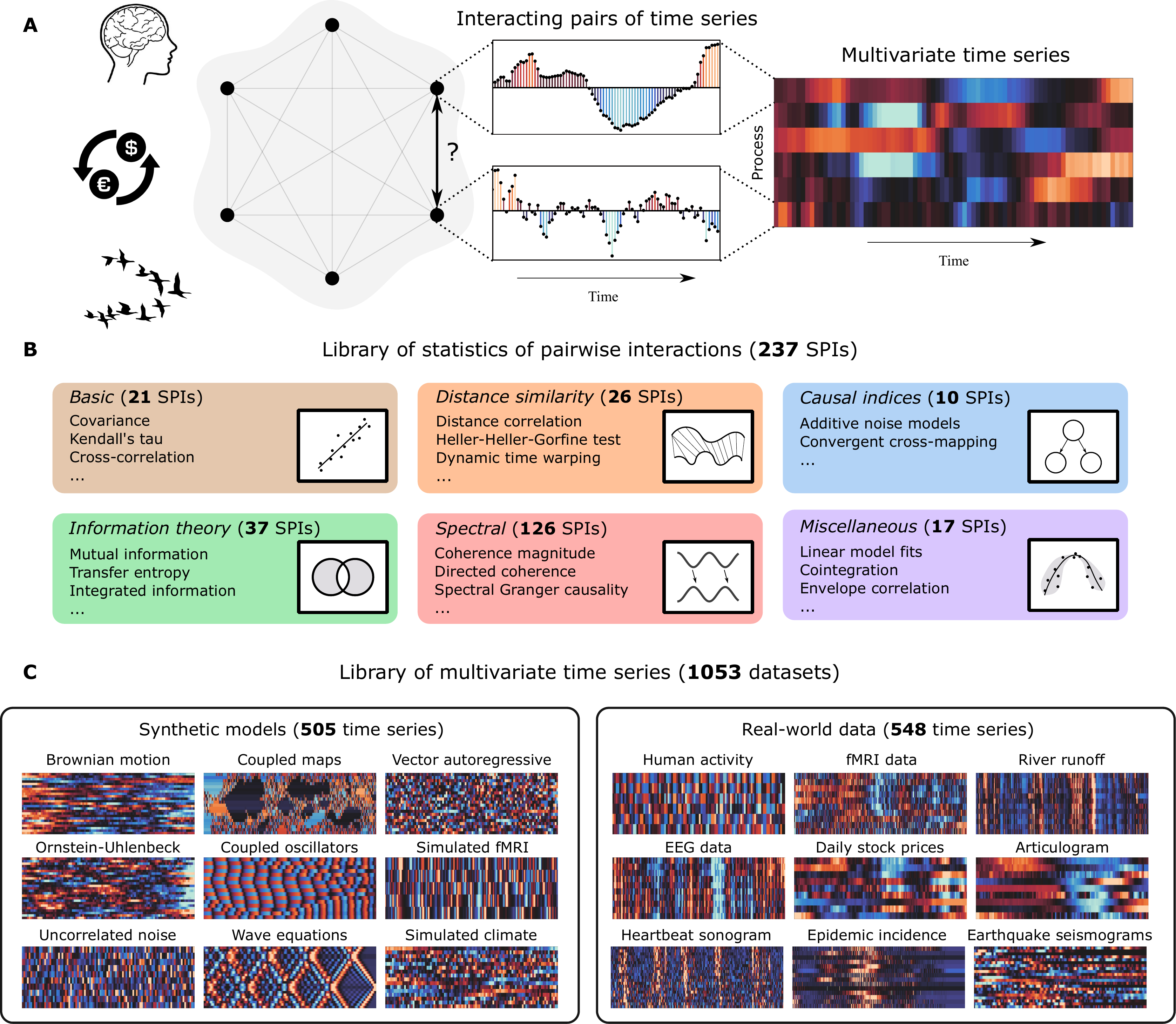}
    \caption{
    \textbf{We developed a scientific library comprising 237 statistics of pairwise interactions~(SPIs) and evaluated their behavior on another library of 1053 multivariate time series~(MTS).}
    \textbf{A}. In many disciplines, including neuroscience, economics, and biology, scientists analyze interactions between pairs of processes in a multivariate time series (MTS).
    MTS data can be visualized as a heat map, where color indicates how individual processes (rows) evolve over time (columns).
    \textbf{B}. The 237 SPIs analyzed here derive from six broad disciplinary categories: `basic', `distance similarity', `causal indices', `information theory', `spectral', and `miscellaneous', as described in detail in Sec.~S1; selected examples of SPIs within each category are listed here, along with a schematic depiction.
    \textbf{C}. Selected examples from our diverse collection of 1053 MTS, which are used to evaluate the empirical behavior of the SPIs, are plotted as heat maps.
    Our library includes data generated from synthetic models (left) and measured from real-world systems (right), as described in detail in Sec.~S2.
    }
    \label{fig:schematic}
\end{figure}

As illustrated schematically in Fig.~\ref{fig:schematic}A, the common goal of these studies is to extract meaningful pairwise relationships from multivariate time series~(MTS): sets of observations taken regularly over time~\cite{reinsel_elements_2003}.
In the age of big data, the scientific problems that are studied in diverse disciplinary contexts---from genomics to astronomy~\cite{stephens2015big}---require novel ways to extract information from MTS data.
However, from the myriad ways to quantify a statistical dependency between two time series, it remains common practice to manually select a single method with minimal comparison to alternatives.
For instance, Pearson correlation remains the most commonly used tool for measuring pairwise relationships in neuroimaging~\cite{van_den_heuvel_exploring_2010} and Earth systems science~\cite{runge_inferring_2019}, despite rather restrictive (and often unsatisfied) assumptions that the data are serially independent and normally distributed~\cite{cliff2021assessing}.
Fortunately, many sophisticated and powerful algorithms have been developed to overcome the limitations of simple correlation coefficients, including where dependencies are lagged in time (e.g., cross-correlation~\cite{reinsel_elements_2003}), may be misaligned (e.g., dynamic time warping~\cite{sakoe1978dynamic}), or where the knowledge of one variable improves the predictability of another (e.g., Granger causality~\cite{granger1969investigating}).

In this work, we represent algorithms for measuring interactions between pairs of time series as real-valued summary statistics: \textit{statistics of pairwise interactions}~(SPIs).
Figure~\ref{fig:schematic}B illustrates the diverse theoretical tools and types of interactions covered in the scientific literature on SPIs, from covariance (the foundation of statistics and machine learning) to convergent cross-mapping~\cite{sugihara_detecting_2012} (a modern algorithm developed for inferring causal effect in complex ecosystems).
However, because the theory underlying statistical interactions has been developed largely independently across disciplinary contexts, these methods remain disconnected from one another.
In this work, we unify this wealth of interdisciplinary scientific knowledge, empirically connecting previously disjoint methodological traditions and yielding a unified set of tools for quantifying interactions in complex dynamics.

As diverse scientific methods for quantifying pairwise interactions have never been unified, there remain many unanswered questions.
Are all of these methods capturing unique information about the interactions occurring within a system?
Is there synergy between complementary approaches that, when combined, tells us more about the underlying system than any single method can?
Or, do interesting redundancies exist, hinting at some theoretical underpinning that could stimulate new theory and understanding of the techniques used across disciplines?
Following previous highly comparative studies of univariate time series~\cite{fulcher_highly_2013} and graphs~\cite{peach_hcga_2021}, this work addresses these questions by simultaneously evaluating hundreds of different SPIs directly from data.
Our empirical approach involves first assembling the most comprehensive library of methods for measuring pairwise interactions (Fig.~\ref{fig:schematic}B), and then applying them to a large and multidisciplinary library of MTS, illustrated in Fig.~\ref{fig:schematic}C, which we have curated to represent a wide variety of complex dynamics studied across scientific disciplines.


\paragraph*{Comprehensive scientific libraries of methods (SPIs) and data (MTS).}

We constructed a comprehensive, annotated library of 237 SPIs that we organized into six broad categories based on their underlying theory:
`basic' (e.g., covariance, Kendall's $\tau$~\cite{kendall1938new}, and cross-correlation);
`distance similarity' (e.g., distance correlations~\cite{shen2020distance}, kernel-based independence tests~\cite{gretton2007kernel,heller2013consistent}, and dynamic time warping~\cite{sakoe1978dynamic});
`causal indices' (e.g., additive noise models and convergent cross-mapping~\cite{sugihara_detecting_2012});
`information theory' (e.g., Granger causality~\cite{granger1969investigating,barnett_granger_2009},
 transfer entropy~\cite{schreiber_measuring_2000}, and integrated information~\cite{oizumi2016measuring});
`spectral' (derived from Fourier or wavelet transformations, e.g., coherence magnitude, phase-locking value~\cite{lachaux1999measuring}, and spectral Granger causality~\cite{dhamala_estimating_2008});
and `miscellaneous' (e.g., cointegration~\cite{engle1987co} and model fits).
A full list of SPIs, along with descriptions and references is in Sec.~S1.
Accompanying this paper is an extendable python-based toolkit, \emph{pyspi}~\cite{cliff2021pyspi}, that includes implementations of all SPIs and allows users to leverage a unified interdisciplinary literature by extracting hundreds of measures of interaction from any MTS.

To understand the behavior of these SPIs on data, we constructed a library of 1053 diverse model-generated and real-world MTS, with the aim of capturing the main classes of systems and dynamics that are studied across scientific disciplines, including synchronization, spatiotemporal chaos, wave propagation, criticality, and phase transitions.
Our library contains 505 synthetic MTS generated from mathematical models, including:
uncorrelated and correlated noise (e.g., Cauchy and normally distributed noise, and Brownian motion);
coupled maps (e.g., vector autoregression~\cite{reinsel_elements_2003} and coupled map lattices~\cite{kaneko2011complex});
coupled ordinary differential equations (e.g., Kuramoto oscillators~\cite{strogatz2000kuramoto}, Hodgkin--Huxley and Wilson--Cowan networks);
and partial differential equations (namely, wave equations, in which processes are embedded in physical space).
It also contains 548 diverse real-world MTS from public databases across: geophysics (e.g., earthquake seismograms and atmospheric processes);
medicine (e.g., heartbeat sonograms, functional magnetic resonance imaging (fMRI) data, and electroencephalograms (EEGs));
physiology (e.g., accelerometer and gyroscope readings for sports and basic motions);
and finance (e.g., exchange rates and stock prices), among others.
Each MTS comprises between 5--40 processes and between 100--2000 observations, characteristics that were designed to match many real-world datasets.
Across all MTS in our library, we have a total of 195\,112 pairwise interactions that we used to evaluate the SPIs.
Descriptions of all MTS are in Sec.~S2, and the full database accompanies this article~\cite{cliff2021database}, enabling scientists to test their methods on a diverse sample of complex dynamics.

\subsection*{Organizing pairwise interactions by their empirical behavior}

\begin{figure}[t!]
    \centering
    \includegraphics[width=\textwidth]{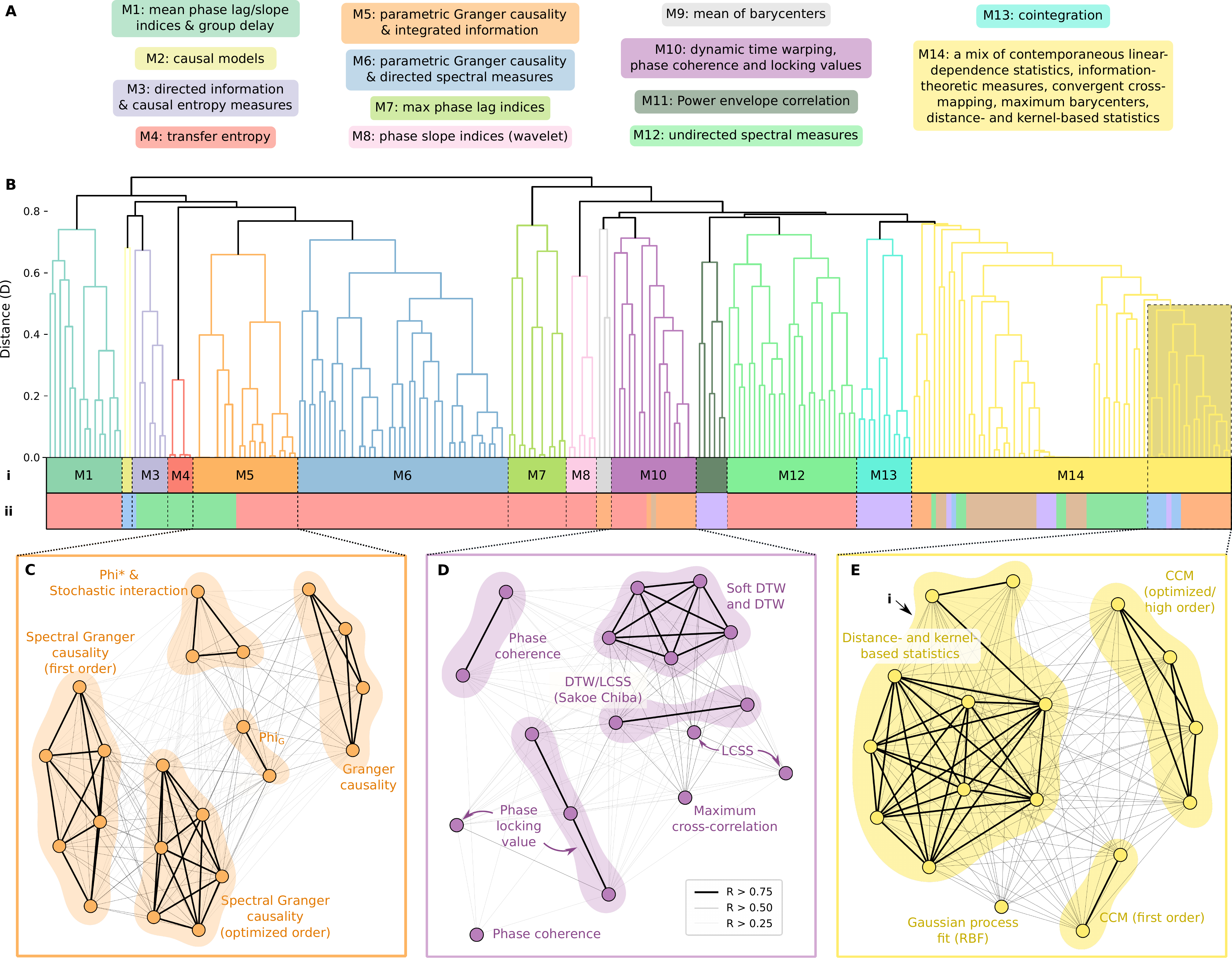}
    \caption{
    \textbf{Statistics for measuring pairwise interactions (SPIs) between time series can be organized into 14 modules based on their behavior on over 1000 multivariate time series (MTS), providing an intuitive, data-driven organization of an interdisciplinary scientific literature.}
    \textbf{A}. A brief summary of the main types of methods in each of the modules (see Sec.~S1 for a full list with descriptions).
    \textbf{B}. The dendrogram used to infer the modules, produced by hierarchical clustering using a dissimilarity, $D = 1 - R$, based on the empirical similarity index, $R$ (see \emph{Methods}).
    SPIs are colored according to: (\textbf{i}) their module label (upper row), and (\textbf{ii}) their literature categorization from Fig.~\ref{fig:schematic}B (lower row).
    A high-resolution version of this dendrogram, including each of the SPIs (leaf nodes) that form the modules, is in Fig.~\ref{S-fig:full_dendrogram}.
    Three selected modules that include a mixture of SPIs developed in different disciplinary contexts are shown as network plots in: \textbf{C}. M5, \textbf{D}. M10, and \textbf{E}. A subset of M14.
    In these plots, SPIs are represented as nodes, three different edge weight thresholds are displayed (corresponding to $R > 0.25$, $R > 0.5$, and $R > 0.75$), and connected components with high similarity, $R > 0.75$, are shaded.
    Abbreviations used in \textbf{C}--\textbf{E}: DTW (dynamic time warping), LCSS (longest continuous subsequence), CCM (convergent cross-mapping), and RBF (radial basis function).
    }
    \label{fig:dendrogram}
\end{figure}

Having assembled our libraries of methods (as 237 SPIs) and data (as 1053 MTS), we aimed to analyze how similarly the different SPIs behave on the data.
To achieve this, we developed an empirical similarity index, $R$ ($0 \leq R \leq 1$), that captures the relationship between any two SPIs by comparing their output when applied to all 195\,112 pairwise interactions present in the 1053 MTS.
As detailed in \textit{Methods}, our index is derived from the average absolute Spearman correlation between a pair of SPIs when applied to all pairs of processes in all datasets.
The minimum value, $R = 0$, indicates a pair of maximally distinct SPIs (with uncorrelated behavior on all datasets), while the maximum, $R = 1$, indicates a pair of SPIs that are perfectly correlated on all datasets (i.e., behave as simple monotonic transformations of one another).
As $R$ captures the average across all datasets, a pair of SPIs with high $R$ reflects a broad similarity of behavior across MTS containing very different types of structure, and therefore acts as a suitable candidate index of empirical similarity.

Using the dissimilarity measure, $D = 1 - R$, we organized all 237 SPIs using hierarchical clustering, yielding the dendrogram shown in Fig.~\ref{fig:dendrogram}B.
This provides a data-driven, structured representation of a diverse literature that allows us to probe and interpret relationships between scientific methods at multiple levels.
We focus our analysis here on a 14-module resolution (with modules labeled M1--M14), which captures important methodological connections between groups of SPIs with similar behavior on diverse MTS.
As summarized in Fig.~\ref{fig:dendrogram}A, these fourteen modules grouped common conceptual and theoretical approaches to measuring interactions between pairs of time series, demonstrating the ability of our empirical approach to meaningfully organize the interdisciplinary literature.

In addition to grouping similar types of methods into modules, we found that different high-level conceptual formulations of dynamical interactions were recapitulated in the relationships between modules.
For example, Modules M3--M6 contain distinct types of SPIs---including Granger causality~\cite{granger1969investigating}, directed information~\cite{massey1990causality}, and integrated information~\cite{tononi2004information,oizumi2016measuring}---which all capture statistical dependencies between two time series by considering the context of their past.
This idea that observable interactions are predicated (or, from a statistical standpoint, conditioned) on the history of a process was first proposed by the Wiener--Granger theories of `causality' and `feedback'~\cite{wiener1956theory}, specifically by measuring how one time series might improve the self-predictability of another.
Our results group SPIs based on this underlying theoretical formulation, due to their characteristic behavior on data.
Other types of SPIs (that do not predicate on the self-predictability of a process) also display distinctive behavior, including measures of contemporaneous relationships (the correlation coefficients of M14), dependencies that account for temporal lags (the coherence measures of M12), or temporal dilation and shifts (dynamic time warping and related methods in M10).

Ten of the fourteen modules are `homogeneous', containing methods that are derived from similar underlying theory, indicated by the color of the category labels in Fig.~\ref{fig:dendrogram}B(ii).
Of these ten homogeneous modules, six of them (M4, M7, M8, M9, M11, and M13) comprise SPIs for measuring one specific type of pairwise interaction, only differing in either their algorithms (e.g., M13 contains both the Engle--Granger~\cite{engle1987co} and the Johansen~\cite{pesaran1999autoregressive} tests for measuring cointegration), summary statistics (e.g., M8 contains both mean and maximum of the wavelet-based phase lag index~\cite{nolte_robustly_2008}), or parameter settings (e.g., M4 contains five estimation techniques for transfer entropy~\cite{schreiber_measuring_2000}).
The remaining four homogeneous modules (M1, M2, M3, and M12) comprise methods with very similar theoretical underpinnings, e.g., M12 contains many SPIs for measuring undirected interactions via Fourier transformations, such as the magnitude and the imaginary part of the coherence~\cite{bastos_tutorial_2016}.
Of particular interest are the four `heterogeneous' modules (M3, M5, M10, and M14), which mix SPIs from different literature categories, revealing interesting connections between different theoretical bases for quantifying pairwise dependence.
While M3 contains a mix of SPIs based on information theory (six labeled `information-theoretic measures' and one labeled as a `causal index': information-geometric conditional independence), the remaining three modules establish interesting connections between the behavior of seemingly disparate SPIs on MTS data.
Three networks that are derived from these modules are plotted in Figs~\ref{fig:dendrogram}C--E and are investigated in detail below.

Module M5, shown in Fig.~\ref{fig:dendrogram}C, contains an intriguing mix of two distinct types of methods: (i) five linear estimators for integrated information: $\Phi_G$~\cite{oizumi2016unified}, $\Phi^*$~\cite{oizumi2016measuring}, and stochastic interaction~\cite{ay2015information}; and (ii) 16 estimators for Granger causality, in both the time and frequency domains~\cite{geweke_measurement_1982}.
While Granger causality and integrated information theory were developed in very different contexts---Granger's investigations into `causality' between economic time series in 1969~\cite{granger1969investigating} versus Tononi's recent integrated information theory ($\Phi$) of consciousness~\cite{tononi2004information,oizumi2016measuring,oizumi2016unified}---our analysis reveals that all SPIs in this module nevertheless behave similarly on real data (with an average empirical similarity, $\langle R \rangle = 0.52$, in the 95th percentile of all $R$ values, see Fig.~S1C).
Despite their distinct disciplinary contexts, recent results have indeed shown that Granger causality can be formulated as the information-theoretic measure transfer entropy~\cite{barnett_granger_2009}, and can thus be grouped under the same information-geometric framework as integrated information theory~\cite{oizumi2016unified,cliff_minimising_2018}.
However, it was not known whether or not these information-theoretic SPIs behave similarly in practice, and as such, their relationship is not widely recognized.
Module M5 thus demonstrates an important confirmation of our empirical approach in being able to recapitulate emerging theory and unify scientific tools for understanding interacting processes.

Module M10, shown in Fig.~\ref{fig:dendrogram}D, highlights striking connections between three conceptually distinct types of methods:
(i) dynamic time warping (DTW), which was developed in the data-mining community to quantify the similarity between two (potentially shifted and dilated) audio signals~\cite{sakoe1978dynamic};
(ii) cross-spectral phase-based measures---the maximum phase coherence~\cite{bastos_tutorial_2016} and the mean and maximum phase-locking value~(PLV)~\cite{lachaux1999measuring}---which were developed to examine frequency-specific synchronization in neuroimaging data~\cite{bastos_tutorial_2016};
and (iii) the maximum cross correlation~\cite{reinsel_elements_2003}, a classic statistical technique for correlating two time series at different lags.
All of these SPIs capture time-lagged interactions between two processes, but in slightly different ways: the maximum cross-correlation finds the highest fixed-lag match, DTW extends this idea by optimizing the distortion of the time axis to best match potentially misaligned time series, and the cross-spectral measures account for time lags in terms of phase differences.
Module M10 thus reveals new connections between diverse approaches to capturing associations between pairs of potentially unaligned time series, indicating a common conceptual basis for methods developed and applied across disciplines---whether they are measuring synchronization between neuro-electric recordings or recognizing speech from audio signals.

Finally we discuss module M14, which groups 66 SPIs from all literature categories except for `spectral' (see Fig.~\ref{fig:schematic} for categories).
In part, this module recapitulates certain theoretical relationships that have already been established, such as the equivalence between linear-Gaussian mutual information and absolute correlation~\cite{mackay_information_2003} (with a maximum similarity of $R = 1$).
To highlight some new relationships, we focus on a demonstrative submodule, shown in Fig.~\ref{fig:dendrogram}E, which comprises 17 SPIs from the `causal indices', `distance similarity', and `miscellaneous' literature categories.
We first note the tight cluster of SPIs, labeled `i' in Fig.~\ref{fig:dendrogram}E, which were developed independently in two different domains: distance correlation-based methods~\cite{shen2020distance} from the statistics community, and kernel-based methods from the machine-learning community.
This cluster first highlights a recent finding that distance correlation and the Hilbert--Schmidt Independence Criteria (HSIC, a kernel-based method)~\cite{gretton2007kernel} are equivalent when computed using certain distance kernels~\cite{sejdinovic2013equivalence}; our results suggest that similar theoretical connections can be established between the other SPIs of the cluster (including the Heller--Heller--Gorfine test~\cite{heller2013consistent} and multi-scale graph correlation~\cite{shen2020distance}).
Second, we find that the distance-based and kernel-based statistics display strikingly similar behavior as common implementations of the convergent cross-mapping (CCM) algorithm, which was originally developed for inferring causality in complex ecosystems~\cite{sugihara_detecting_2012}.
CCM measures the causal effect of one time series on another by the ability of the second to reconstruct the first with a nearest-neighbor approach.
The fact that these methods behave so similarly on MTS data indicates that the well-studied techniques of phase-space reconstruction (used in the CCM algorithm) have a correspondence to the nonlinear kernel-estimation techniques from the statistics and machine learning communities.
These observed connections also have important practical ramifications, e.g., our results suggest candidate proxy algorithms to substitute for the computationally expensive CCM, which could yield major computational efficiencies and enable new applications on larger datasets.

\subsection*{Leveraging diverse methods to address scientific problems}

\begin{figure}[h!]
    \centering
    \includegraphics[width=0.95\textwidth]{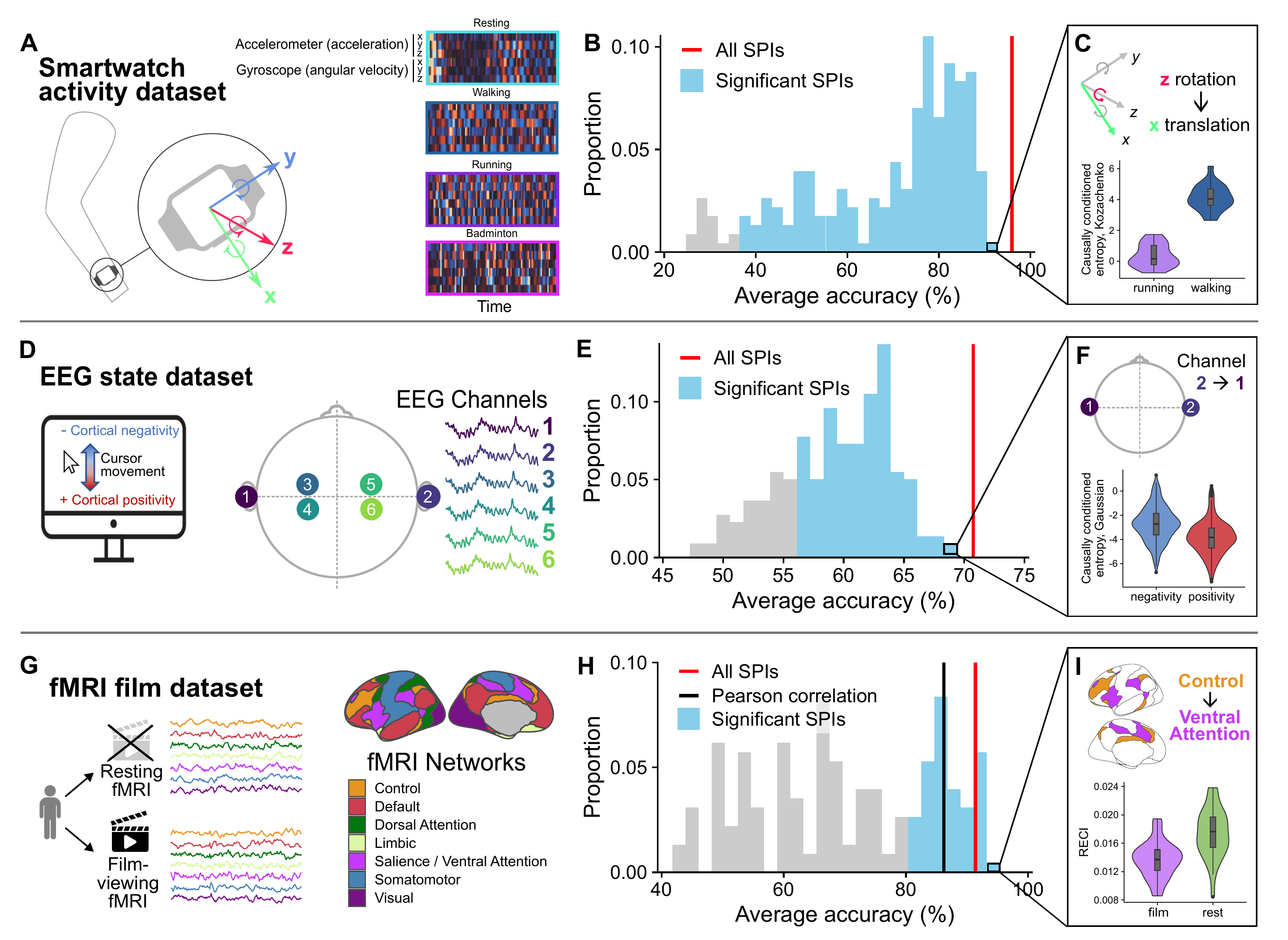}
    \caption{
    \textbf{A comprehensive library of SPIs can be used to accurately classify and understand differences in human movement and neural activity datasets.}
    \textbf{A}. In the smartwatch activity dataset, we aimed to determine which of four activities a participant is performing (`resting', `walking', `running', or `badminton') from smartwatch accelerometer recordings.
    An example MTS from each of the four classes is shown as a heat map.
    \textbf{B}. Distribution of average classification accuracy (over train--test resamples) across 228 SPIs, where all pairwise interactions are used as the basis for classification for each SPI.
    The 213 SPIs with significant accuracy are shaded blue (permutation test, Bonferroni-corrected $p < 0.05$).
    Concatenating feature vectors from all SPIs into a single classification model yields an average accuracy of 96\% (red).
    \textbf{C}. A violin plot showing the distribution of the top-performing SPI---causally conditioned entropy with a Kozachenko--Leonenko density estimator---between the wrist's $x$-axis translation and $z$-axis rotation, indicating a stronger interaction during walking (blue) than running (purple).
    \textbf{D}. In the EEG state dataset we aimed to classify positive versus negative cortical activity states using a six-channel EEG dataset.
    \textbf{E}. Distribution of average classification accuracy across all 219 SPIs, including the 165 significant SPIs (shaded blue), and combination of all SPIs (71\% accuracy, red).
    \textbf{F}. The top-performing SPI, a causally conditioned entropy, is visualized from EEG Channels 2 to 1 in cortical negativity (blue) versus positivity (red).
    \textbf{G}. In the fMRI film dataset, we aimed to classify rest versus film-viewing using fMRI time series from seven brain networks.
    \textbf{H}. Distribution of classification accuracy across all 227 SPIs, including the 67 significant SPIs (shaded blue) and annotating the performance of Pearson correlation coefficient (86\%, black) and the combined set of all 227 SPIs (91\%, red) for comparison.
    \textbf{I}. The top-performing SPI, regression error-based causal inference (RECI), is visualized from the control to ventral attention networks in film-viewing (purple) and rest (green) conditions.
    }
    \label{fig:classification}
\end{figure}

Our results above illustrate the rich diversity of scientific methods for quantifying pairwise interactions.
This diversity suggests that, when quantifying pairwise interactions for a given application, there is potential to compare across the scientific literature of SPIs to:
(i) select the best-performing SPI in an unbiased, data-driven way; and
(ii) leverage a synergistic combination of multiple complementary SPIs to better capture complex underlying interactions in MTS.
Here we provide a simple demonstration of this highly comparative strategy to three MTS classification problems using three open datasets:
(1) \textit{Smartwatch activity} dataset (Fig.~\ref{fig:classification}A), where the aim is to classify one of four behavioral states (walking, running, resting, or playing badminton) from six-sensor smartwatch MTS (a 3-axis accelerometer and 3-axis gyroscope) \cite{ruiz2021great};
(2) \textit{EEG state} dataset (Fig.~\ref{fig:classification}D), where the aim is to distinguish positive versus negative slow cortical potentials from single-subject electroencephalogram (EEG) data (originally used to move a cursor up or down on a computer screen) \cite{bagnall2018uea, birbaumer1999spelling}; and
(3) \textit{fMRI film} dataset (Fig.~\ref{fig:classification}G), where the aim is to classify resting and film-watching conditions from functional magnetic resonance imaging (fMRI) data \cite{byrge2019high,byrge2020accurate}.
To investigate the performance of different SPIs on these tasks, we represented each MTS as a set of features corresponding to all pairwise interactions between its constituent processes and compared their classification performance using a linear SVM with cross-validation (see \textit{Methods} for details).

We observed a wide range of SPI performance in each case study, ranging from null performance up to high and significant performance:
27--92\% accuracy on the smartwatch activity dataset (Fig.~\ref{fig:classification}B);
48--69\% accuracy on the EEG state dataset (Fig.~\ref{fig:classification}E); and
41--95\% accuracy on the fMRI film dataset (Fig.~\ref{fig:classification}H).
Many SPIs displayed significant classification performance on each dataset (permutation test, Bonferroni-corrected $p < 0.05$): 213 SPIs (smartwatch activity), 165 SPIs (EEG state), and 67 SPIs (fMRI film).
This wide range of observed SPI performance on all three datasets demonstrates the crucial importance of selecting an SPI that is able to capture the relevant types of interactions underlying a given dataset.

To understand the types of interactions that characterize the labeled classes of MTS, we analyzed and interpreted the highest-performing SPIs on each dataset (all results are provided in Table S3).
To provide a simple demonstration of this process, here we focus on the top-performing individual SPI in each case study.
In the smartwatch activity dataset, the top-performing SPI was causally conditioned entropy (CCE) using a Kozachenko--Leonenko estimator (92\% accuracy; SPI label \texttt{cce\_kozachenko}, see Sec.~S1.4.7 for details).
Its high performance is driven in part by its ability to capture the coupling from the wrist's $z$-rotation to its $x$-translation, which differs strongly between running and walking, as shown in Fig.~\ref{fig:classification}C.
This tells us that elbow movement in the $x$-direction is more informative of subsequent $z$-rotation (as the wrist rotates outward in the $z$-direction) in walking than in running.

The top-performing SPI for the EEG state dataset was, as above, a causally conditioned entropy (CCE), but using a Gaussian density estimator (69\%; SPI label \texttt{cce\_gaussian}, see Sec.~S1.4.7 for details).
Its performance was driven in part by its ability to capture the increased coupling from EEG Channel 2 to 1 (from near the right ear to near the left ear) in cortical negativity versus positivity states, as shown in Fig.~\ref{fig:classification}F.
Other statistics designed to capture directed information flow (in a way that includes instantaneous interactions in the presence of feedback) also performed well on this task, including directed information with a Gaussian density estimator (69\% accuracy; SPI label \texttt{di\_gaussian}, see Sec.~S1.4.8).
While prior applications of SPIs derived from directed information theory are limited \citep{liu2012quantification, mehta2017directional}, our highly comparative analysis suggests them as high-performing methods for measuring EEG coupling alongside other novel candidates for further investigation, including the direct directed transfer function evaluated over high frequencies (67\%; SPI label \texttt{ddtf\_multitaper\_mean\_fs-1\_fmin-0-25\_fmax-0-5}, see Sec.~S1.5.9) and the Hilbert--Schmidt Independence Criterion (66\%; SPI label \texttt{hsic}, see Sec.~S1.2.5).
Of the classical methods for quantifying EEG connectivity, some are recapitulated as high performers by our data-driven analysis, including mean directed coherence across various frequency bands (all 67\% accuracy; see Sec.~1.5.10; cf. \citep{wang1992directed}), while others exhibited surprisingly low accuracy, such as algorithmic variants of partial directed coherence \citep{schelter2006testing} (between 55\% and 64\%).

Finally, for the fMRI film dataset, the top-performing SPI was regression error-based causal inference (95\% accuracy; SPI label \texttt{reci}, see Sec~S1.3.4 for details).
Its high performance is driven in part by its ability to capture the stronger coupling from the control network to the ventral attention network during film-watching compared to rest (Fig.~\ref{fig:classification}I).
The dominant way of measuring coupling in whole-brain fMRI is to use the Pearson correlation coefficient \cite{van_den_heuvel_exploring_2010, Smith2011:NetworkModellingMethods} (annotated in Fig.~\ref{fig:classification}H), and it exhibits strong and statistically significant classification accuracy on this problem (86\%).
However, our data-driven approach highlights 30 alternative SPIs with higher performance (88\% to 95\%, listed in Table~S3).
These high-performing methods include alternative types of covariance (e.g., minimum covariance determinant, 93\% accuracy; SPI label \texttt{cov\_MinCovDet}, see Sec.~S1.1.1) and precision estimates (e.g., using Ledoit--Wolf shrinkage, 91\%; SPI label \texttt{prec\_LedoitWolf}, see Sec.~S1.1.2) that better deal with non-Gaussian bivariate distributions.
Other high performers include information theoretic SPIs (like conditional entropy, joint entropy, and mutual information using Gaussian density estimators, all 91\% accuracy, see Sec.~1.4.1--1.4.3) and directed SPIs which distinguish asymmetric coupling (e.g., the top-performing regression error-based causal inference, \texttt{reci}).
Compared to the typically subjective process of selecting an appropriate method to analyze a given dataset, the highly comparative approach demonstrated here highlights the most useful scientific methods automatically, facilitating interpretable understanding of the conceptual formulations of pairwise dependence that drive successful performance.

Since different types of systems involve different types of interactions between measured processes, we expected different SPIs to perform well across the three datasets.
Indeed, we found that an SPI with high performance on one problem does not imply its high performance on other problems.
For example, some SPIs performed well only on a single dataset, like dynamic time warping with an Itakura parallelogram (SPI label \texttt{dtw\_constraint-itakura}, see Sec.~S1.2.7) which was a top performer on the fMRI film dataset (91\%), but yielded weaker performance on the smartwatch activity dataset (78\%), and null performance on the EEG state dataset (53\%).
But some SPIs did perform strongly across all three datasets, such as the cross distance correlation (SPI label \texttt{dcorrx\_maxlag-10}, see Sec.~S1.2.2), which ranked among the top 10 SPIs for all three problems (90\%, 66\%, and 91\%, for the smartwatch activity, EEG state, and fMRI film datasets, respectively).
Moreover, different algorithmic variants of causally conditioned entropy were top performers for the smartwatch activity (\texttt{cce\_kozachenko}, 92\%), EEG state (\texttt{cce\_gaussian}, 69\%), and fMRI film (\texttt{cce\_kernel\_W-0.5}, 90\%) datasets (although we note a strong dependence of the density estimation approach on CCE performance).
We also found that grouping SPIs based on the fourteen data-driven modules (identified by their similarity of behavior on real data, Fig.~\ref{fig:dendrogram}) better captured their relative performance on these tasks than the six literature categories (Fig.~\ref{fig:schematic}), as shown in Supplementary Fig.~\ref{S-fig:literature_vs_module}, suggesting our modular representation as a useful one for understanding differential SPI performance on a given task.

Relative to investigating individual SPIs one at a time, we finally aimed to investigate the value of drawing on multiple SPIs simultaneously.
We developed a combined representation of the pairwise dependence structures captured by all SPIs, allowing us to simultaneously represent each MTS using a large and diverse set of pairwise dependency structures (through feature concatenation, cf. \textit{Methods}).
Although this approach represents each MTS in a much higher-dimensional space than the individual-SPI representation analyzed above (with associated challenges for robust classifier fitting), we expected it to outperform the individual best SPI on datasets involving multiple types of interactions, such that simultaneously leveraging multiple SPIs provides complementary and useful information about class differences.
Relative to the top-performing individual SPI, our combined-SPI approach improved classification accuracy on the smartwatch activity dataset (to 96\%) and the EEG state dataset (to 71\%), shown as vertical red lines in Figs~\ref{fig:classification}B and E.
On the fMRI film dataset, it yielded slightly lower performance (91\%) than the top individual SPI, \texttt{reci} (95\%), suggesting that the associated interactions are well-captured by a single, well-chosen SPI on this dataset (i.e., multiple SPIs do not provide an advantage sufficient to overcome the challenges of fitting a classifier in a higher-dimensional space).
By simultaneously drawing on a wide range of SPIs, the simple statistical approach demonstrated here can quantify multiple complementary types of interactions from MTS data (and is likely to yield improved accuracy through optimization, see \textit{Discussion}).

\subsection*{Discussion}

Despite its rather restrictive assumptions, Pearson correlation has remained the analytical standard for measuring pairwise interactions in time-varying systems for over 100 years~\cite{runge_inferring_2019, van_den_heuvel_exploring_2010, sugihara_detecting_2012}.
But as scientists have continued to record and analyze data from increasingly complex systems (such as the brain, climate, and financial markets), the use of contemporaneous linear correlations as a default way of measuring relationships between time series has been challenged.
Here we took an empirical approach to unifying a large and disjoint literature of sophisticated methods for characterizing pairwise interactions in time-varying systems by studying the behavior of 237 SPIs on a library of 1053 diverse MTS, and across three specific applications.
Our findings demonstrate the power of our highly comparative empirical approach to studying SPIs by:
(i) grouping methods that rely on similar underlying theory, highlighting previously reported and new theoretical connections between diverse methods;
(ii) capturing common, higher-level conceptualizations of dynamical interactions (such as whether methods predicate a statistical dependence on self-predictability); and
(iii) automatically uncovering high-performing and interpretable methods tailored to the types of interactions underlying accelerometer, EEG, and fMRI time series.

Understanding relationships between disparate theoretical literatures is crucial for connecting ideas across the sciences.
Our results reveal a rich variety of algorithmic approaches to quantifying interactions from time series, providing a unified understanding of the most useful methods scientists have developed to date.
We showed how methods for quantifying pairwise interactions organize around common underlying theoretical formulations of dynamical dependency, such as how contemporaneous associations are treated by conditioning on the past of time series (following the Wiener--Granger conceptualization of `causality' and `feedback'~\cite{granger1969investigating}).
Our empirical approach also recapitulates known theoretical findings (such as the relationship between Granger causality and transfer entropy \cite{barnett_granger_2009}) and highlights new connections between methods developed in different theoretical traditions, thereby highlighting fruitful directions for future research to consolidate, extend, and develop new theory.
The results also have practical benefits, including the ability to select computationally simpler and more clearly interpretable algorithms that nevertheless display similar behavior as complex and computationally intensive methods (e.g., for CCM, Fig.~\ref{fig:dendrogram}E).


As different time-varying systems contain different statistical relationships between their elements---e.g., that capture instantaneous or time-delayed responses, linear or nonlinear interactions, condition on the past, allow for non-constant delays, or infer directional coupling---our highly comparative approach can detect the most informative types of SPIs for capturing the relevant types of interactions underlying a given dataset.
Unlike many machine-learning approaches to MTS classification, which are challenging to interpret~\cite{ruiz2021great}, the highly comparative approach connects scientists to interpretable theory that shapes understanding of the most important types of pairwise interactions in a dataset.
It follows recent approaches to wide comparison across interdisciplinary literatures on summary statistics of univariate time series \cite{fulcher_highly_2013, Fulcher2018:FeaturebasedTimeseriesAnalysis, Fulcher2017:HctsaComputationalFramework} and complex networks \cite{peach_hcga_2021}.
This broad, comprehensive methodological comparison stands in contrast to a more conventional approach in which the data analyst manually selects a method, a practice that leaves open the possibility that alternative methods may provide clearer interpretation, better performance, or computational efficiencies.
The three classification case studies analyzed here provide a simple demonstration of the procedure, automatically highlighting high-performing SPIs from across the literature and providing interpretable understanding of the relevant interactions in each dataset.
We observed a wide range of performance in all cases, highlighting the importance of careful selection of SPIs for a given application.
In the EEG state dataset, for example, our analysis flagged high-performing SPIs consistent with common methodological practice in EEG analysis (thus recapitulating existing  domain knowledge) and others with surprisingly poor performance, as well as flagging novel high-performing SPIs as promising candidates for methodological innovation in the field.
Future work investigating which types of SPIs are best suited to which types of problems will yield new insights into the interactions underlying different complex systems, and is likely to uncover additional novel applications of SPIs to new problems.

Since the behavior of an SPI depends on the types of interactions present in a dataset, our empirical organization of SPIs shown in Fig.~\ref{fig:dendrogram} is dependent on our library of MTS.
As a simple illustrative example, consider the case that our MTS library only contained data generated by contemporaneously linearly coupled processes, e.g., lag-1 vector autoregressive models, VAR(1).
Then the unique behavior of more sophisticated SPIs (such as those capturing time-lagged or nonlinear dependence) would not be observed, because data containing those types of interactions would not be present.
Consequently, the variety of behaviors exhibited by our library of SPIs would be reduced, and the modular structure evident in the dendrogram of Fig.~\ref{fig:dendrogram} would become far more homogeneous.
In constructing the MTS data library used in this work, we have thus aimed to be as even-handed and comprehensive as possible in sampling diverse MTS, which has been sufficient to yield a meaningful and useful representation of the interdisciplinary literature on SPIs.
However, as any such finite sample is incomplete and relies on subjective decisions in its construction, future work may explore the dataset-dependent similarity of SPI pairs in detail to construct more nuanced organizations of the literature.
This would provide new understanding of how different mechanisms (and hence empirical dependency structures) play out in different classes of complex systems.
Future work may also revisit the simple methodological choices made here, including our decisions to:
(i) quantify the similarity of a pair of SPIs using a single number ($R$);
(ii) represent the resulting relationships in one dimension (as a dendrogram);
and (iii) analyze the resulting dendrogram at a single selected resolution (a 14-module decomposition).
For example, more complex unsupervised methods, including overlapping community-detection methods \cite{Evans2009:LineGraphsLink, Ahn2010:LinkCommunitiesReveal}, could reveal interesting new relationships between SPIs at different scales of resolution.

While the ability to compare across the full library of SPIs is powerful, it comes at a computational cost, particularly for larger MTS datasets in which each MTS is sampled over many time points and processes.
The high redundancy between many groups of SPIs (Fig.~\ref{fig:dendrogram}), and preliminary indications from the case studies that some SPIs may exhibit generally high performance, suggests the potential for major efficiency gains by constructing a high-performing and minimally redundant reduced set of SPIs.
Beyond a simple approach of selecting a representative SPI from each of the 14 modules, a more systematic approach could compare SPI performance across a wide range of representative MTS tasks, e.g., following recent work reducing a large pool of thousands of interdependent univariate time-series features to just 22 representatives \cite{Lubba2019:Catch22CAnonicalTimeseriesa}.
There is also much scope for statistical optimization in applying any given set of SPIs to real-world problems, beyond the simple choices made here.
In particular, performance could be optimized by tackling the challenges of high-dimensionality and strong inter-dependence of SPIs using dimensionality reduction, regularization, or feature-selection methods to find small but highly explanatory combinations of complementary SPIs tailored to a given dataset \cite{Phinyomark2017:NavigatingFeaturesTopologically}.
Given the methodological freedoms involved in highly comparative time-series analysis, the registered report style should be considered, in which a subset of the data is analyzed in Stage 1, and results on the remaining data are reported in Stage 2 (e.g., \cite{rowe2023neuralinput, leung_psyarxiv_2022}).

Beyond the applications to MTS classification presented here, the ability to represent the coupling structure in a MTS using a large and diverse set of pairwise interactions could form the foundation for tackling myriad statistical inference problems, including regression, time-series clustering, anomaly detection, and causal network inference.
For example, current methods for inferring networks of causal interactions from MTS data \cite{Peixoto2019:NetworkReconstructionCommunity, Banerjee2023:NetworkInferenceShort} range from heuristic approaches based on thresholding a set of pairwise dependencies defined by a given SPI, through to full statistical inference \cite{Peixoto2019:NetworkReconstructionCommunity, Hoffmann2020:CommunityDetectionNetworks, McCabe2021:NetrdLibraryNetwork, Wollstadt2019:IDTxlInformationDynamics}.
Rather than manually selecting an SPI for this purpose, the ability to compare diverse SPIs provides new flexibility to capture different types of underlying interactions that could form the basis for new, more flexible network-inference algorithms.
Such an extension may also explore extending beyond point estimates of the strength of pairwise dependencies (as considered here), towards assessing statistical significance through comparison to an appropriate null model \cite{cliff2021assessing, Peel2022:StatisticalInferenceLinks, Banerjee2023:NetworkInferenceShort}.
Future work could also explore comprehensive statistical representations of MTS using more diverse properties than just the set of pairwise interactions, by also incorporating properties of univariate dynamics of individual system components \cite{Fulcher2017:HctsaComputationalFramework}, system-level structure in the full set of pairwise interactions represented as a network \cite{peach_hcga_2021}, and higher-order interactions \cite{Battiston2021:PhysicsHigherorderInteractions}.

The flexible and extendable software accompanying this work, \emph{pyspi}~\cite{cliff2021pyspi}, allows scientists to adopt our highly comparative approach to pairwise interactions in their own applications, including comparing across diverse SPIs to extract the best-performing scientific methods for a given task.
Its application, e.g., through careful framing of systems-level inference \cite{Peel2022:StatisticalInferenceLinks}, has the potential to highlight unexpected and powerful methodological approaches to quantifying interaction patterns in time-varying systems.
The large data library of MTS provided with this work~\cite{cliff2021database} is a resource that will allow researchers to characterize the behavior of their computational methods on a comprehensive range of real-world and simulated systems, addressing issues associated with only testing new complex systems methods on idealized datasets \cite{Peel2022:StatisticalInferenceLinks}.
This work demonstrates the utility of an empirical approach to unifying diverse complex dynamical systems and their methods of analysis, providing new insights and tools for scientific discovery.

\section*{Methods}

\subsection*{The empirical similarity index}
\label{sec:similarity-stat}

Our main aim in this work was to assess the similarity of behavior of any two SPIs across a diverse range of MTS.
In this section we describe the empirical similarity index, $R$, used to capture the similarity between the behavior of all pairs of SPIs across our diverse library of 1053 MTS.
This index measures the monotonic relationship between a pair of SPIs across a large number of datasets that have different numbers of processes (and thus a different number of pairwise interactions).
For an $M$-variate time series, $\bm{z} = (z_1,\ldots,z_M)$, we first consider the $M \times M$ matrix of pairwise interactions (MPI), constructed by evaluating a given SPI for all pairs of $M$ processes within the MTS.
The scalar values of the MPI, $S = (s_{ij}) \in \mathbb{R}^{M \times M}$, where the $(i,j)$ entry of the matrix, $s_{ij} = s(z_i,z_j)$, denotes an SPI evaluated on the $i$th and the $j$th time series, $z_i$ and $z_j$.
In general, undirected statistics (for which $s_{ij} = s_{ji}$, such as Kendall's $\tau$) yield symmetric MPIs, whereas directed statistics (which have $s_{ij} \neq s_{ji}$, such as Transfer Entropy) yield asymmetric MPIs.
Furthermore, some SPIs are signed (e.g., correlation coefficients are within $[-1,1]$), while others are unsigned (e.g., distance correlation is within $[0,1]$).
To compare all methods appropriately, we converted signed SPIs to their absolute value, such that all statistics increase with the strength of dependency between $z_i$ and $z_j$ (e.g., we analyze the magnitude of the covariance rather than its sign).
Some example MPIs are shown in Fig.~\ref{S-fig:MPI_schematic}A.

Having computed MPIs for all SPIs on all 1053 MTS, we computed the empirical similarity index between each pair of SPIs via a two-step process (depicted in Fig.~\ref{S-fig:MPI_schematic}B).
First, we defined the similarity of a pair of SPIs, $k$ and $l$, on a given dataset, $d$.
We did this by computing MPIs for $k$ and $l$ (yielding $S_{kd}$ and $S_{ld}$, respectively) and then computing the absolute value of the Spearman's rank correlation coefficient between each of their (off-diagonal) entries, $|r_d|$.
The resulting correlation value, $|r_d|$, thus captures the strength of a monotonic relationship between the output of the two SPIs on the dataset $d$.
After repeating this computation for all MTS in our library, we calculated the empirical similarity index, $R$, as the average of $|r_d|$ across all datasets.

While $R$ provides a useful scalar summary of the similarity between a pair of SPIs, it is important to note that some pairs of SPIs have quite a wide distribution of scores, $|r_d|$, across datasets, indicating that they yield highly correlated outputs on some MTS (high $|r_d|$), but not on others (low $|r_d|$).
For more details, including a detailed visual breakdown of $|r_d|$ distributions for some selected pairs of SPIs, see Sec.~\ref{S-sec:similarity-stat}.
We found a wide range of similarity indices across pairs of SPIs, from $R \approx 0$ to $R \approx 1$, illustrated by the cumulative distribution function in Fig.~\ref{S-fig:MPI_schematic}C.

Note that our main results are not highly dependent on the choice to summarize the distribution of $|r_d|$ scores as a mean rather than a median.
The two choices yield a set of SPI--SPI similarity scores (that form the basis of the clustering shown here) that are highly correlated to each other, $\rho = 0.99$, and result in a modular decomposition that is highly overlapping (a Rand index of 0.95 \cite{Rand1971}, using the same threshold, $R = 0.76$, to form clusters).


\subsection*{Classification case study}
\label{sec:classification-case-study}

\subsubsection*{Datasets}
\label{sec:classification-datasets}
The smartwatch activity dataset is derived from the \texttt{BasicMotions} problem in the University of East Anglia (UEA) MTS classification repository \cite{bagnall2018uea}.
Each MTS includes six sensors---a 3-axis accelerometer and a 3-axis gyroscope---recorded for 10\,s at 10\,Hz, yielding 1000 time points.
There are 20 MTS in each of four classes (resting, walking, running, or playing badminton), for a total of 80 MTS in the dataset.
This dataset was recently analyzed in a large MTS classification challenge~\cite{ruiz2021great} in which the data was split 50/50 into training and test subsets with 30 stratified repeats.
The baseline classifier (based on dynamic time warping) achieved 95.25\% accuracy; the best algorithm (HIVE--COTE) achieved 100\% accuracy \cite{ruiz2021great}.

The EEG state dataset corresponds to the \texttt{SelfRegulationSCP1} problem in the UEA MTS classification repository \cite{bagnall2018uea} and was originally published in \citet{birbaumer1999spelling}.
Electrical activity was measured from six EEG channels in one participant as they were instructed to move a cursor up or down on a computer screen by generating negative or positive slow cortical potentials, respectively.
The physical placement of these EEG electrodes is depicted in Fig.~\ref{fig:classification}D.
Cortical activity (measured in $\mu$V) was recorded for 3.5\,s at 256\,Hz, yielding 896 time points in each channel per MTS, with 282 negativity trials and 279 positivity trials.
This dataset was analyzed in \citet{ruiz2021great}, assessing accuracy using 30 stratified train--test splits of 268 training and 293 test samples.
In the original comparison, the baseline DTW algorithm achieved an accuracy of 81.81\% and the best algorithm (TapNet) achieved 95.68\% \cite{ruiz2021great}.

The fMRI film dataset is derived from a functional connectome fingerprinting study examining individual signatures of cortical activity in $N = 29$ individuals at rest or while watching a film \cite{byrge2019high, byrge2020accurate}.
In this dataset, blood-oxygen-level-dependent (BOLD) signals were recorded in 114 parcellated cortical regions with a repetition time (TR) of 813\,ms (a sampling rate of 1.23\,Hz) for either 1200 frames (resting) or between 952--1000 frames (film-watching).
Each trial type (rest versus film-watching) was repeated four times per participant.
Here we analyzed pre-processed data obtained from \citet{Betzel2020}.
For consistency and simplicity, we examined the first rest and first film-watching session per participant; across all participants and trials, only the first $947$ frames contain real data, so we restricted our analysis to these time ranges.
To retain a comparable number of processes as the first two classification case studies, we averaged BOLD signals from the 114 original brain regions into the seven functional networks from \citet{yeo2011organization}, as depicted in Fig.~\ref{fig:classification}G.
We compare the performance of the Pearson correlation coefficient, used to construct functional connectivity matrices in the original publications \cite{byrge2019high,byrge2020accurate}, to our library of SPIs.

\subsubsection*{Classification}
For all three case studies, our simple approach to SPI-based classification involved computing the matrix of pairwise interactions (MPI)---$6 \times 6$ for the smartwatch activity and EEG state datasets, and $7 \times 7$ for the fMRI film dataset---for each $z$-scored MTS and repeating for each SPI.
We then used the elements of these matrices as features for a linear support vector machine (SVM) classifier.
Note that we used the most recent version of \textit{pyspi} (v0.4.0) to compute SPIs for the classification case studies, which included some improved implementations of some SPIs.
Features were extracted from each MPI differently for directed and undirected SPIs (see Table S1): for undirected SPIs (for which the corresponding MPI is symmetric), we used the upper triangular entries as features, whereas for directed SPIs, we used all non-diagonal elements as features.
As a preprocessing step, for each case study, we removed any SPI that had invalid entries (due to numerical issues) in any of the MPIs, or gave constant results across all MTS (omitted SPIs are listed in Table S2).
This yielded a set of $228$ SPIs for the smartwatch activity problem, $219$ SPIs for the EEG state problem, and $227$ SPIs for the fMRI film problem.
For the analysis involving combining all SPIs into a single classifier, this yielded a total of $4755$ features for the smartwatch activity dataset, $4659$ features for the EEG state dataset, and $6743$ features for the fMRI film dataset.

The linear SVM was implemented using default settings from Python's \texttt{\detokenize{scikit-learn}}~\cite{pedregosa2011scikit} as part of a classification pipeline that involved $z$-score feature normalization (fitted on training data and applied to unseen test data).
The very simple methodological choices made here allowed us to focus on demonstrating the key conceptual types of analyses made possible by drawing on a diverse set of SPIs, aiding transparency while acknowledging that more complicated statistical methodologies are likely to improve the classification performance quoted here.
We implemented 30 class-stratified train--test splits for cross-validation with the same proportions implemented in \cite{ruiz2021great} using the \texttt{StratifiedShuffleSplit} function for the smartwatch activity and EEG state datasets.
Since there are $N = 29$ individuals in the fMRI film dataset, we implemented leave-one-individual-out cross-validation, such that each classifier was trained with the rest and film scans of $N = 28$ participants and tested on the rest and film fMRI scans of the left-out participant.

We measured classification performance using total accuracy for all three case studies.
Statistical significance was estimated using a permutation testing approach whereby 100 null models were fitted (using randomly shuffled class labels) and evaluated using the same cross-validation classification procedure described above.
The observed classification performance for each SPI was then compared with the pooled null distribution of all SPIs \cite{leek2011joint}, yielding $p$-values that were then adjusted for multiple comparisons by controlling the family-wise error at 0.05 using the method of Bonferroni \cite{bonferroni1936teoria}.
The performance metric for the union of all SPIs was similarly compared with its corresponding null permutation distribution to yield a single $p$-value per classification problem.

\section*{Data availability statement}

The $1053$ MTS datasets generated and analyzed in this study are available online~\cite{cliff2021database}.
All data used in the case studies are from open sources, as described above in \textit{Methods} \cite{bagnall2018uea, Betzel2020}.

\section*{Code availability statement}

In addition to the open-source toolkit that accompanies this paper~\cite{cliff2021pyspi}, all code for reproducing the results for the empirical similarity index methods and figures of this paper are available online~\cite{cliff2022pyspi-paper}.
All code for reproducing the results for the classification case studies are also available online \citep{annie_g_bryant_2023_8027702}.

\section*{Acknowledgments}

OMC, NT and BDF were supported by NHMRC Ideas Grant 1183280.
AGB was supported by an Australian Government Research Training Program (RTP) Scholarship, an American Australian Association Graduate Education Fund Scholarship, and the University of Sydney Physics Foundation.
High performance computing facilities provided by the School of Physics, University of Sydney contributed to our results.

\bibliography{main}
\bibliographystyle{benbibstyle}

\end{document}


\title{Supplementary Text for `Unifying Pairwise Interactions in Complex Dynamics'}
\author[1,2]{Oliver M. Cliff}
\author[1,2]{Annie G. Bryant}
\author[2,3]{Joseph T. Lizier}
\author[4,5,6]{Naotsugu Tsuchiya}
\author[1,2]{Ben D. Fulcher}
\affil[1]{School of Physics, The University of Sydney, Camperdown NSW 2006, Australia.}
\affil[2]{Centre for Complex Systems, The University of Sydney, Camperdown NSW 2006, Australia.}
\affil[3]{Faculty of Engineering, The University of Sydney, Camperdown NSW 2006, Australia.}
\affil[4]{Turner Institute for Brain and Mental Health \& School of Psychological Sciences, Faculty of Medicine, Nursing, and Health Sciences, Monash University, Melbourne, Victoria, Australia.}
\affil[5]{Center for Information and Neural Networks (CiNet), National Institute of Information and Communications Technology (NICT), Suita-shi, Osaka 565-0871, Japan.}
\affil[6]{Advanced Telecommunications Research Computational Neuroscience Laboratories, 2-2-2 Hikaridai, Seika-cho, Soraku-gun, Kyoto 619-0288, Japan.}

\date{\vspace{-5ex}}

\maketitle

This supplementary document includes detailed information about the statistics for pairwise interactions (SPIs) in Sec.~\ref{sec:spis-library} and the data included in our multivariate time series~(MTS) library in Sec.~\ref{sec:MTS-library}.
We also provide additional information about the empirical similarity index in Sec.~\ref{sec:similarity-stat}.

\section{Library of statistics for pairwise interactions (SPIs)}
\label{sec:spis-library}

This section introduces our library of 237 SPIs that we have collected for evaluating pairwise interactions in multivariate time series (MTS).
Each method is a real-valued measurement, $s(x,y) \in \mathbb{R}$, of some type of dependency between two time series, $x$ and $y$, and as such are referred to as statistics for pairwise interactions (SPIs).
We only include methods that can be computed on continuous-valued $M$-variate time series (see Sec.~\ref{sec:MTS-library}), which can be used to compute an $M$-by-$M$ square matrix of pairwise interactions (referred to as an MPI, see Sec.~\ref{sec:similarity-stat}).
Prior to evaluating each statistic, we standardize ($z$-score) the MTS along the time axis.
We include methods that both operate directly on pairs of time series (i.e., bivariate time series, like Kendall's $\tau$~\cite{kendall1938new}) and methods that operate directly on the full MTS (like precision matrices); both are used to generate an MPI from an MTS.

Most SPIs in our library are based on existing implementations, but in some cases, especially where the output of a method was not a single real number, we implemented new real-valued statistics.
Moreover, many of the algorithms include a number of free parameters that we set either using optimization procedures (often available within implementations) or fix to a small number of sensible predefined settings.
The combination of both the different parameter configurations and the different summary statistics taken from each method gives each SPI in the library a unique identifier (as a string) that will be used throughout this supplement.
Consider the SPI identifier `\code{xcorr_mean_sig-True}' (cf. Sec.~\ref{sec:xcorr}); here, \code{xcorr} refers to the method of cross-correlation between $x$ and $y$, which itself does not provide a single statistic but rather a correlogram (i.e., a series of correlation coefficients for each time delay between the two signals).
However, the two additional modifiers that are separated by underscores in the identifier, \code{mean} and \code{sig-True}, gives a scalar value.
The first modifier, \code{mean}, indicates that we are taking the average across lags of the cross-correlation function.
The second modifier, \code{sig-True}, indicates that we will only take the mean over statistically significant lags.
Therefore, the combination of the method (\code{xcorr}) with the modifiers (\code{mean} and \code{sig-True}) gives a scalar value from the bivariate time series $(x,y)$.
By using different parameters and modifiers of 45 distinct methodologies, we obtain the 237 SPIs used throughout this work.

In order to simplify our presentation of the methods (and their modifiers) that are used to generate these SPIs, we have organized each method into one of the following broad literature categories: `basic', `distance similarity', `causal', `information theory', `spectral', and `miscellaneous'.
Each of these categories is included as a subsection below.
Note that this grouping is for convenience and is not unambiguous---as we find throughout this work, many existing statistics fit multiple descriptions and could thus be organized validly into multiple categories.
Finally, for each SPI, we include a small number of keywords to indicate whether the interactions they measure are: undirected or directed; linear or nonlinear; signed or unsigned; bivariate or multivariate; and contemporaneous, time-dependent, frequency-dependent, or time-frequency dependent.
These keywords help users to understand key assumptions of each of the SPIs.
Note that our aim for the \textit{pyspi} library is for it to be an evolving library that is refined and improved over time, with this section capturing an early snapshot of it.

\subsection{Basic statistics (21 SPIs)}

In this section, we detail the 21 SPIs that we have categorized as `basic statistics'.

\subsubsection{Covariance (\code{cov})}
\textit{Keywords: undirected, linear, signed, multivariate, contemporaneous.}

The covariance matrix is estimated for a wide variety of statistical procedures.
Due to the $z$-scoring of the time series, the correlation and covariance matrices are equivalent and thus the covariance statistic is within $[-1,1]$.
We use v0.24.1 of \textit{scikit-learn}~\cite{pedregosa2011scikit} to compute the covariance matrix via a number of estimators: the standard maximum likelihood estimate (MLE) (denoted by the modifier \code{EmpiricalCovariance}); the elliptic envelope (\code{EllipticEnvelope})~\cite{rousseeuw1999fast} and the minimum covariance determinant (\code{MinCovDet}) methods for outlier removal; the Lasso technique, which uses an $\ell_1$-regularization to sparsify the covariance matrix \code{GraphicalLasso} (as well as a method with the regularization method chosen through cross validation with five splits, \code{GraphicalLassoCV}); a basic shrinkage covariance estimator with a fixed shrinkage coefficient of 0.1 (\code{ShrunkCovariance})~\cite{pedregosa2011scikit}; the Ledoit-Wolf method for optimizing the shrinkage coefficient (\code{LedoitWolf})~\cite{ledoit2004well}; and the oracle approximating shrinkage, an improved method for optimizing the shrinkage coefficient if the data are Gaussian (\code{OAS})~\cite{chen2010shrinkage}.
By using all estimators, we obtain 8 SPIs:
\begin{enumerate}[noitemsep]
    \begin{multicols}{2}
        \item \code{cov_EmpiricalCovariance}
        \item \code{cov_EllipticEnvelope}
        \item \code{cov_MinCovDet}
        \item \code{cov_GraphicalLasso}
        \item \code{cov_GraphicalLassoCV}
        \item \code{cov_ShrunkCovariance}
        \item \code{cov_LedoitWolf}
        \item \code{cov_OAS}
    \end{multicols}
\end{enumerate}

\subsubsection{Precision (\code{prec})}
\textit{Keywords: undirected, linear, signed, multivariate, contemporaneous.}

The precision matrix is the matrix inverse of the covariance matrix, and can be used to quantify the association between each pair of time series while controlling for concomitant effects of all other time series.
For normalized time-series data, the precision matrix is equivalent to the partial correlation between each pairwise time series, conditioned on all other time series, and is within $[-1,1]$.
The precision matrix is computed via the same module as the covariance matrix (in \textit{scikit-learn}~\cite{pedregosa2011scikit}), and has the same estimators.
By using all estimators, we obtain 8 SPIs:
\begin{enumerate}[resume,noitemsep]
    \begin{multicols}{2}
        \item \code{prec_EmpiricalCovariance}
        \item \code{prec_EllipticEnvelope}
        \item \code{prec_MinCovDet}
        \item \code{prec_GraphicalLasso}
        \item \code{prec_GraphicalLassoCV}
        \item \code{prec_LedoitWolf}
        \item \code{prec_OAS}
        \item \code{prec_ShrunkCovariance}
    \end{multicols}
\end{enumerate}

\subsubsection{Spearman's rank-correlation coefficient (\code{spearmanr})}
\textit{Keywords: undirected, nonlinear, signed, bivariate, contemporaneous.}

Spearman's $\rho$ is a nonparametric measure of rank correlation between variables.
The use of ordinal (ranked) variables allows the statistic to capture non-linear (but monotonic) relationships between random variables.
The method is implemented via function \code{spearmanr} in \textit{SciPy}~\cite{2020SciPy-NMeth}, and has a value in $[-1,1]$.
\begin{enumerate}[resume,noitemsep]
    \item \code{spearmanr}
\end{enumerate}

\subsubsection{Kendall's rank-correlation coefficient (\code{kendalltau})}
\textit{Keywords: undirected, nonlinear, signed, bivariate, contemporaneous.}

Kendall's $\tau$~\cite{kendall1938new} assesses the association of ordinal variables, similar to Spearman's $\rho$, but has certain differences, such as becoming more mathematically tractable in the event of ties~\cite{gilpin1993table}.
The method is implemented via function \code{kendalltau} in \textit{SciPy}~\cite{2020SciPy-NMeth}, and has a value in $[-1,1]$.
\begin{enumerate}[resume,noitemsep]
    \item \code{kendalltau} \label{spi:kendalltau}
\end{enumerate}

\subsubsection{Cross correlation (\code{xcorr})}
\textit{Keywords: undirected, linear, signed/unsigned, bivariate, time-dependent.}

\label{sec:xcorr}

The cross-correlation function is defined as the Pearson correlation between two time series for all lags~\cite{brockwell2009time}, giving values in $[-1,1]$ for each lag.
To estimate the cross-correlation function, we use \textit{SciPy} \cite{2020SciPy-NMeth}, which outputs a correlogram, i.e., the correlation from the MLE of the cross-covariance at a given lag, normalised by the autocovariance.
The cross-correlation is computed with fewer observations at larger lags and so it is common to truncate the function at a given level~\cite{afyouni2019effective,cliff2021assessing}, which we do by only using the first $T/4$ lags, where $T$ is the number of observations.
Moreover, a correlation below $1.96/\sqrt{T}$ is considered statistically insignificant~\cite{brockwell2009time}, thus we optionally cut-off the lags at this level (the modifier \code{sig-True} means we only use the statistically significant values, and \code{sig-False} means we use all values).
We take the two summary statistics of the correlogram: the maximum over the considered lags (denoted by modifier \code{max}), and the average over the considered lags (\code{mean}).
By using each approach outlined above, we obtain three SPIs:

\begin{enumerate}[resume,noitemsep]
    \begin{multicols}{2}
        \item \code{xcorr_max_sig-True}
        \item \code{xcorr_mean_sig-True}
        \item \code{xcorr_mean_sig-False}
    \end{multicols}
\end{enumerate}

\subsection{Distance similarity (26 SPIs)}

In this section, we detail SPIs that we have categorized as `distance-based similarity' measures in that they aim to establish statistical similarity or independence based on the pairwise distance between bivariate observations.
All of the methods presented in this section are implemented using one of the following toolboxes:
\begin{itemize}
    \item For distance correlation (and related) statistics, as well as reproducing kernel Hilbert space (RKHS)-based statistics, we use the \textit{Hypothesis Testing in Python (hyppo)} package~\cite{panda2019hyppo}.
    \item For dynamic time warping (and related) statistics, we use the \textit{tslearn} package~\cite{tavenard2020tslearn}.
\end{itemize}

\subsubsection{Distance correlation (\code{dcorr})}
\textit{Keywords: undirected, nonlinear, unsigned, bivariate, contemporaneous.}
\label{sec:dcorr}

Distance correlation is used to infer the independence between two random variables via pairwise distance metrics and hypothesis tests~\cite{szekely2007measuring}.
The sample distance correlation is computed by summing over the entry-wise product of Euclidean distance matrices.
This statistic is biased, however an unbiased estimator can be obtained by first double-centering the distance matrices~\cite{shen2020distance}.
Although any pairwise distance metric can be used, here we only use Euclidean distance because distance correlation computed with this metric has been shown to be universally consistent (asymptotically converging to the true value).
We compute both the biased and unbiased statistics from the \textit{hyppo} package, yielding two SPIs:
\begin{enumerate}[resume,noitemsep]
    \begin{multicols}{2}
        \item \code{dcorr}
        \item \code{dcorr_biased}
    \end{multicols}
\end{enumerate}

\subsubsection{Cross distance correlation (\code{dcorrx})} \label{spi:dcorrx}
\textit{Keywords: undirected, nonlinear, unsigned, bivariate, time-dependent.}

The cross-distance correlation~\cite{panda2019hyppo} quantifies the independence between two univariate time series based on distance correlation~\cite{szekely2007measuring} (see above, Sec.~\ref{sec:dcorr}).
This measure is the average of lagged distance correlations between the past of time series $x$ to the future of time series $y$, up to a given lag.
We include both a low-order (lag-1, \code{maxlag-1}) and a high-order (lag-10, \code{maxlag-10}) assumption of the number of relevant lags, yielding two SPIs:
\begin{enumerate}[resume,noitemsep]
    \begin{multicols}{2}
        \item \code{dcorrx_maxlag-1}
        \item \code{dcorrx_maxlag-10}
    \end{multicols}
\end{enumerate}

\subsubsection{Multiscale graph correlation (\code{mgc})}
\textit{Keywords: undirected, nonlinear, unsigned, bivariate, contemporaneous.}
\label{sec:mgc}

Multiscale graph correlation (MGC)~\cite{shen2020distance} is a generalization of distance correlation~\cite{szekely2007measuring} (see Sec.~\ref{sec:dcorr}) that is designed to overcome its limitations in inferring nonlinear relationships such as circles and parabolas.
Specifically, the algorithm truncates the Euclidean distance matrices (the procedure at the core of distance correlation) at an optimal threshold.
MGC also includes a smoothing function that is intended to remove any bias introduced by disconnected components in the graph introduced by truncating the distance matrix.
Only this unbiased estimator is included here (as in \code{hyppo}):
\begin{enumerate}[resume,noitemsep]
    \item \code{mgc}
\end{enumerate}

\subsubsection{Cross multiscale graph correlation (\code{mgcx})}
\textit{Keywords: undirected, nonlinear, unsigned, bivariate, time-dependent.}

The cross-multiscale graph correlation~\cite{panda2019hyppo} is defined similarly to \code{dcorrx} (Sec.~\ref{spi:dcorrx}) but uses lagged MGCs instead of lagged distance correlations (see Sec.~\ref{sec:mgc}).
By taking lag-1 (\code{maxlag-1}) and lag-10 (\code{maxlag-10}), we get two SPIs:
\begin{enumerate}[resume,noitemsep]
    \begin{multicols}{2}
        \item \code{mgcx_maxlag-1}
        \item \code{mgcx_maxlag-10}
    \end{multicols}
\end{enumerate}

\subsubsection{Hilbert-Schmidt Independence Criterion (\code{hsic})}
\textit{Keywords: undirected, nonlinear, unsigned, bivariate, contemporaneous.}
\label{sec:hsic}

The Hibert-Schmidt Independence Criterion (HSIC)~\cite{gretton2007kernel} is an RKHS-based statistic that quantifies a statistical dependence between random variables via a sample kernel matrix.
As with distance correlation, HSIC yields a biased statistic, where the unbiased (and consistent) estimator can be derived by double centering the kernel distance matrix~\cite{gretton2007kernel}.
Both biased and unbiased estimators are computed using \textit{hyppo}, yielding two SPIs:
\begin{enumerate}[resume,noitemsep]
    \begin{multicols}{2}
        \item \code{hsic}
        \item \code{hsic_biased}
    \end{multicols}
\end{enumerate}

\subsubsection{Heller-Heller-Gorfine Independence Criterion (\code{hhg})}
\textit{Keywords: undirected, nonlinear, unsigned, bivariate, contemporaneous.}

The Heller-Heller-Gorfine~(HHG) method~\cite{heller2013consistent} yields an RKHS-based statistic that uses the ranks of random variables to obtain sample kernel matrices, rather than their distances (cf. HSIC in Sec.~\ref{sec:hsic}).
This SPI is computed via the \textit{hyppo} package:
\begin{enumerate}[resume,noitemsep]
    \item \code{hhg}
\end{enumerate}

\subsubsection{Dynamic time warping (\code{dtw})}
\textit{Keywords: undirected, nonlinear, unsigned, bivariate, time-dependent.}

Dynamic time warping (DTW)~\cite{sakoe1978dynamic} extends the ideas of measuring the pairwise Euclidean distance between time series by allowing for potentially dilated time series of variable size.
Specifically, DTW finds the minimum distance between two time series through alignment (shifting and dilating of the sequences).
This algorithm (and many outlined below) also include the Sakoe-Chiba band~\cite{sakoe1978dynamic} and the Itakura parallelogram~\cite{itakura1975minimum} global constraints on the alignments to prevent pathological warpings.
We compute this statistic using the \textit{tslearn} package for the three global constraints (i.e., no constraints (no modifier), the band (modifier \code{sakoe_chiba}), and the parallelogram (\code{itakura})), yielding three SPIs:
\begin{enumerate}[resume,noitemsep]
    \begin{multicols}{2}
        \item \code{dtw}
        \item \code{dtw_constraint-itakura}
        \item \code{dtw_constraint-sakoe_chiba}
    \end{multicols}
\end{enumerate}

\subsubsection{Longest common subsequence (\code{lcss})}
\textit{Keywords: undirected, nonlinear, unsigned, bivariate, time-dependent.}

The longest common subsequence~\cite{vlachos2002discovering} generalizes ideas from DTW by measuring the similarity between \textit{continuous subsections} of time series, rather than the entire time series themselves, subject to distance thresholds and alignment constraints.
We use the default threshold from the \textit{tslearn} package ($\epsilon=1$), and include the three global constraints discussed above:
\begin{enumerate}[resume,noitemsep]
    \begin{multicols}{2}
        \item \code{lcss}
        \item \code{lcss_constraint-itakura}
        \item \code{lcss_constraint-sakoe_chiba}
    \end{multicols}
\end{enumerate}

\subsubsection{Soft dynamic time warping (\code{softdtw})}
\textit{Keywords: undirected, nonlinear, unsigned, bivariate, time-dependent.}

Soft dynamic time warping~\cite{cuturi2017soft} uses a smoothed formulation of DTW to optimize the minimal-cost alignment as a differentiable loss function.
This method is computed via the \textit{tslearn} package (with the default hyperparameter, $\gamma=1$), and includes the same global constraints as DTW:
\begin{enumerate}[resume,noitemsep]
    \begin{multicols}{2}
        \item \code{softdtw}
        \item \code{softdtw_constraint-itakura}
        \item \code{softdtw_constraint-sakoe_chiba}
    \end{multicols}
\end{enumerate}

\subsubsection{Barycenter (\code{bary})}
\textit{Keywords: undirected, nonlinear, unsigned, bivariate, time-dependent.}

A barycenter (or Fr\'{e}chet mean)~\cite{petitjean2011global,schultz2018nonsmooth} is a (univariate) time series that minimizes the sum of squared distances between MTS.
The \textit{tslearn} package provides functions to obtain barycenters by minimizing the sum-of-squares of the following distance metrics: (unwarped) Euclidean distance (\code{euclidean}); alignment via expectation maximization (\code{dtw}); alignment via a differential loss function (\code{softdtw}); and alignment via a subgradient descent algorithm (\code{sgddtw}).
For each pair of (bivariate) time series in an $M$-variate MTS, we compute the raw barycenters then summarize them by taking their mean (modifier \code{mean}) and maximum (\code{max}), yielding seven SPIs:
\begin{enumerate}[resume,noitemsep]
    \begin{multicols}{2}
        \item \code{bary_euclidean_max}
        \item \code{bary_dtw_mean}
        \item \code{bary_dtw_max}
        \item \code{bary_softdtw_mean}
        \item \code{bary_softdtw_max}
        \item \code{bary_sgddtw_mean}
        \item \code{bary_sgddtw_max}
    \end{multicols}
\end{enumerate}

\subsection{Causal inference (10 SPIs)}

In this section, we detail the 10 SPIs that we have categorized as `causal inference'.
These statistics aim to establish directed independence from bivariate observations, typically by making assumptions about the underlying model.
We use the following two packages:
\begin{itemize}
    \item For convergent cross-mapping, we use the \textit{Empirical Dynamic Modeling (pyEDM)} package~\cite{pyEDM}.
    \item For all other SPIs, we use v0.5.23 of the \textit{Causal Discovery Toolbox (cdt)}~\cite{kalainathan2019causal}.
\end{itemize}

\subsubsection{Additive noise model (\code{anm})}
\textit{Keywords: directed, nonlinear, unsigned, bivariate, contemporaneous.}
\label{sec:anm}

Additive noise models~\cite{hoyer2008nonlinear} are used for hypothesis testing directed nonlinear dependence (or causality) of $x \to y$ by making the assumption that the effect variable, $y$, is a function of a cause variable, $x$, plus a noise term (that is independent of the cause).
In this framework we use the statistic from \textit{cdt} as our SPI, which is computed by first predicting $y$ from $x$ via a Gaussian process (with a radial basis function kernel), and then computing the normalized HSIC test statistic from the residuals (see Sec.~\ref{sec:hsic} above) yielding the SPI:
\begin{enumerate}[resume,noitemsep]
    \item \code{anm}
\end{enumerate}

\subsubsection{Information-geometric causal inference (\code{igci})}
\textit{Keywords: directed, nonlinear, unsigned, bivariate, contemporaneous.}

Information-geometric causal inference~\cite{daniusis2012inferring} is a method for inferring causal influence from $x\to y$ for deterministic systems with invertible functions.
The statistic is computed using \textit{cdt} as the difference in differential entropies where the probability density is computed via nearest-neighbor estimators.
\begin{enumerate}[resume,noitemsep]
    \item \code{igci}
\end{enumerate}

\subsubsection{Conditional distribution similarity fit (\code{cds})}
\textit{Keywords: directed, nonlinear, unsigned, bivariate, contemporaneous.}

The conditional distribution similarity fit~\cite{fonollosa2019conditional} is the standard deviation of the conditional probability distribution of $y$ given $x$, where the distributions are estimated by discretizing the values.
\begin{enumerate}[resume,noitemsep]
    \item \code{cds}
\end{enumerate}

\subsubsection{Regression error-based causal inference (\code{reci})}
\textit{Keywords: directed, nonlinear, unsigned, bivariate, contemporaneous.}

The regression error-based causal inference method~\cite{blobaum2018cause} is an estimate of the causal effect of $x \to y$ by quantifying the error in a regression of $y$ on $x$ with a monomial (power product) model.
In the bivariate case, this statistic is the MSE of the linear regression of the cubic (plus constant) of $x$ with $y$, giving one SPI:
\begin{enumerate}[resume,noitemsep]
    \item \code{reci}
\end{enumerate}

\subsubsection{Convergent cross-mapping (\code{ccm})}
\label{sec:ccm}
\textit{Keywords: directed, nonlinear, unsigned, bivariate, time-dependent.}

The idea behind convergent cross-mapping (CCM)~\cite{sugihara_detecting_2012} is that there is a causal influence from time series $x\to y$ if the Takens time-delay embedding~\cite{takens_detecting_1981} of $y$ can be used to predict the observations of $x$.
The algorithm quantifies the prediction error (in terms of Pearson's $\rho$) of time series $x$ from the delay embedding of time series $y$ for increasing library sizes (i.e., time series length being used in the predictions).
If, as the library size increases, the correlation converges and is higher in one direction than the other, there is an inferred causal link.
The results of CCM are typically represented as two curves~\cite{sugihara_detecting_2012}; one for each causal direction ($x \to y$ and $y \to x$) with the library size on the horizontal axis and the prediction quality (correlation) on the vertical axis.

We use the \textit{pyEDM} package~\cite{pyEDM} to compute CCM, which requires an embedding dimension to be set for the delay embedding of each time series.
We use both fixed embedding dimensions (with dimension 1 and 10, indicated by modifiers \code{E-1} and \code{E-10}, respectively) and an inferred embedding dimension from univariate phase-space reconstruction methods (modifier \code{E-None}).
Following the Supplementary Materials of the original CCM paper~\cite{sugihara_detecting_2012} and the documentation in the \textit{pyEDM} package, we infer the embedding dimension to be the maximum of the two univariate delay embeddings that best predicted each time series.
Given a fixed or inferred embedding dimension, we have an upper and lower bound on the minimum and maximum library size that can be used for computing CCM.
In this work we use 21 uniformly sampled library sizes between this minimum and maximum to generate the CCM curves.
Once the curve (prediction quality as a function of library size) is obtained, we take summary statistics of the mean (modifier \code{mean}) and the maximum (\code{max}) across the curves.
We do not explicitly measure convergence of the algorithm as a function of library size, consistent with common practice in the literature (e.g.,~\cite{pao2021experimentally}), but note that this differs from the original theory; no automatic algorithm or heuristic was originally proposed for quantifying convergence~\cite{sugihara_detecting_2012}.
The combination of statistics and embedding dimension parameters yields six SPIs:
\begin{enumerate}[resume,noitemsep]
    \begin{multicols}{2}
        \item \code{ccm_E-1_mean}
        \item \code{ccm_E-1_max}
        \item \code{ccm_E-10_mean}
        \item \code{ccm_E-10_max}
        \item \code{ccm_E-None_mean} \label{spi:ccm}
        \item \code{ccm_E-None_max}
    \end{multicols}
\end{enumerate}

\subsection{Information theory (37 SPIs)}

The pairwise measures that we employ from information theory are either intended to operate on serially independent observations (e.g., joint entropy and mutual information) or bivariate time series (e.g., transfer entropy and stochastic interaction).
We primarily use v1.5 of the \textit{Java Information Dynamics Toolkit}~\cite{lizier2014jidt} in this section, which allows us to compute differential entropy, mutual information, and transfer entropy, in order to construct many information-theoretic measures.
A density estimation is required to compute information-theoretic measures~\cite{mackay_information_2003}, and in this work we use four different estimators (see references in~\cite{lizier2014jidt}):
\begin{itemize}
    \item The Gaussian-distribution model (denoted by modifier \code{gaussian}) assumes a linear-Gaussian multivariate, where the measure is derived from the cross-covariance matrix;
    \item Kernel estimation (\code{kernel}) uses a box kernel method with a specific kernel width (default width of 0.5 standard deviations (denoted by the modifier \code{W-0.5}));
    \item The Kozachenko-Leonenko technique (\code{kozachenko}) is a nearest-neighbour approach that is suitable when measures can only be constructed using entropy or joint entropy (not mutual information); or
    \item The Kraskov-St\"{o}gbauer-Grassberger~(KSG) technique (\code{ksg}) combines nearest-neighbor estimators for mutual information based measures (default of four nearest neighbors, indicated by modifier \code{NN-4}).
    The KSG estimator is effectively a combination of multiple Kozachenko estimators that includes techniques to remove a bias that is introduced by taking the difference between two differential entropy estimates.
\end{itemize}
An alternative approach to using the continuous estimators above would be to discretise each observation (e.g., through binning) and use a discrete estimator; however, discrete estimators are known to be heavily dependent on the discretization size~\cite{bossomaier2016transfer}.

Density estimates for mutual information (and related measures) are biased by the autocorrelation present in individuals signals~\cite{cliff2021assessing,schreiber_measuring_2000}.
A common solution to reduce bias in nonlinear estimators (like the KSG technique) is to use dynamic correlation exclusion (also known as a Theiler window)~\cite{theiler1986spurious,schreiber_measuring_2000}, which excludes any data points within a given time window from the density estimate and avoids oversampling.
The window size should be large enough to render observations included in the density estimate uncorrelated (sometimes called the ``autocorrelation time''~\cite{theiler1986spurious}); here, we set the window as the product of the autocorrelation functions of both time series (a heuristic for the autocorrelation time based on Bartlett's formula~\cite{cliff2021assessing,bartlett_theoretical_1946}).
The \code{DCE} modifier indicates that the Theiler window is used for the KSG estimator.

\subsubsection{Joint entropy (\code{je})}
\textit{Keywords: undirected, nonlinear, unsigned, bivariate, contemporaneous.}

The joint entropy~\cite{mackay_information_2003} quantifies the uncertainty over the paired observations:
\begin{enumerate}[resume,noitemsep]
    \begin{multicols}{2}
        \item \code{je_gaussian}
        \item \code{je_kozachenko}
        \item \code{je_kernel_W-0.5}
    \end{multicols}
\end{enumerate}

\subsubsection{Conditional entropy (\code{ce})}
\textit{Keywords: undirected, nonlinear, unsigned, bivariate, contemporaneous.}

Conditional entropy~\cite{mackay_information_2003} quantifies the uncertainty over the observations in $y$ in the context of simultaneously observing $x$:
\begin{enumerate}[resume,noitemsep]
    \begin{multicols}{2}
        \item \code{ce_gaussian}
        \item \code{ce_kozachenko}
        \item \code{ce_kernel_W-0.5}
    \end{multicols}
\end{enumerate}

\subsubsection{Mutual information (\code{mi})}
\textit{Keywords: undirected, nonlinear, unsigned, bivariate, contemporaneous.}

Mutual information~\cite{mackay_information_2003} is an undirected measure of the (potentially nonlinear) dependence between paired observations of $x$ and $y$.
\begin{enumerate}[resume,noitemsep]
    \begin{multicols}{2}
        \item \code{mi_gaussian}
        \item \code{mi_kraskov_NN-4}
        \item \code{mi_kraskov_NN-4_DCE}
        \item \code{mi_kernel_W-0.25}
    \end{multicols}
\end{enumerate}

\subsubsection{Time-lagged mutual information (\code{tlmi})}
\textit{Keywords: undirected, nonlinear, unsigned, bivariate, time-dependent.}

Time-lagged mutual information~\cite{schreiber_measuring_2000} is an undirected measure of the dependence between time series $x$ and a time-lagged instance of the series $y$.
We include statistics for only lag-one mutual information, giving four statistics:
\begin{enumerate}[resume,noitemsep]
    \begin{multicols}{2}
        \item \code{tlmi_gaussian}
        \item \code{tlmi_kraskov_NN-4} \label{spi:tlmi}
        \item \code{tlmi_kraskov_NN-4_DCE}
        \item \code{tlmi_kernel_W-0.25}
    \end{multicols}
\end{enumerate}

\subsubsection{Transfer entropy (\code{te})}
\textit{Keywords: directed, nonlinear, unsigned, bivariate, time-dependent.}
\label{sec:te}

Transfer entropy~\cite{schreiber_measuring_2000} is a measure of information transfer from a source time series $x$ to a target time series $y$, based on the Takens time-delay embedding.
Delay embeddings capture the relevant history of a time series that can be used as a predictor of its future and are constructed from an embedding length and a time delay.
The embedding lengths are denoted by modifier's \code{l} and \code{k} for the source, $x$, and target, $y$, respectively; the time delay is denoted \code{lt} and \code{kt} for time series $x$ and $y$.
The embedding parameters can be obtained in a number of ways, so long as their product is (significantly) less than the number of observations.
We compute transfer entropy for both fixed and optimized embedding parameters.
The fixed parameters are minimal values, i.e., typically with a fixed target embedding length of 1 (denoted by \code{k-1}) or 2 (denoted by \code{k-2}).
The optimized parameters are inferred by choosing the embedding parameters (up to a maximum embedding length of 10, denoted by \code{k-max-10}, and maximum time delay of 2, \code{tau-max-2}) that maximize a univariate information-theoretic measure known as active information storage (see the \code{MAX_CORR_AIS} method of~\cite{lizier2014jidt} for details).
Note that there is no inclusion of the \code{gaussian} estimator for transfer entropy, since this is equivalent to Granger causality (see below)~\cite{barnett_granger_2009}.
We also include a symbolic estimator for transfer entropy (denoted by the modifier \code{symbolic}), with details in~\cite{staniek2008symbolic}.
\begin{enumerate}[resume,noitemsep]
    \begin{multicols}{2}
        \item \code{te_kraskov_NN-4_k-max-10_tau-max-4}
        \item \code{te_kraskov_NN-4_DCE_k-max-10_tau-max-4} \label{spi:te}
        \item \code{te_kraskov_NN-4_DCE_k-1_kt-1_l-1_lt-1}
        \item \code{te_kraskov_NN-4_DCE_k-2_kt-1_l-1_lt-1}
        \item \code{te_kraskov_NN-4_k-1_kt-1_l-1_lt-1}
        \item \code{te_kernel_W-0.25_k-1}
        \item \code{te_symbolic_k-1_kt-1_l-1_lt-1}
        \item \code{te_symbolic_k-10_kt-1_l-1_lt-1}
    \end{multicols}
\end{enumerate}

\subsubsection{Granger causality (\code{gc})} \label{sec:gc}
\textit{Keywords: directed, linear, unsigned, bivariate, time-dependent.}

Granger causality~\cite{granger1969investigating} is obtained by assessing directed dependence of $x\to y$ as the predictive power of a bivariate autoregressive model (comprising $x$ and $y$) over the univariate autoregressive model (with $y$ only).
The statistic included in our framework is the $\log$-ratio of residual variance for the two models, which is a variant of Granger causality popularized by Geweke~\cite{geweke_measurement_1982} later found to be equivalent to transfer entropy with a Gaussian estimator~\cite{barnett_granger_2009}.
We compute the time delay and embedding dimension in the same way as transfer entropy (see Sec.~\ref{sec:te}), allowing for both fixed and optimized embedding lengths (\code{k} and \code{l}) and a time delays (\code{kt} and \code{lt}), where the optimization procedure is identical to transfer entropy.
This gives two SPIs:
\begin{enumerate}[resume,noitemsep]
    \begin{multicols}{2}
        \item \code{gc_gaussian_k-max-10_tau-max-2}
        \item \code{gc_gaussian_k-1_kt-1_l-1_lt-1}
    \end{multicols}
\end{enumerate}

\subsubsection{Causally conditioned entropy (\code{cce})}
\textit{Keywords: directed, nonlinear, unsigned, bivariate, time-dependent.}

Causally conditioned entropy~\cite{kramer1998directed,amblard_directed_2011,ziebart2013principle} aims to measure the uncertainty remaining in time series $y$ in the context of the entire (causal) past of both time series $x$ and $y$.
The measure is computed as a sum of conditional entropies (of $y$ given both the past of $x$ and the past of $y$) with increasing history lengths.
The standard assumption is that we consider the entire past of both time series (i.e., there are $T-1$ conditional entropies in the sum, from $1$ to $T-1$); however, for computational reasons, we restrict the history length to $10$.
This corresponds to the assumption that the joint process is, at maximum, a 10th-order Markov chain.
\begin{enumerate}[resume,noitemsep]
    \begin{multicols}{2}
        \item \code{cce_gaussian}
        \item \code{cce_kozachenko}
        \item \code{cce_kernel_W-0.5}
    \end{multicols}
\end{enumerate}

\subsubsection{Directed information (\code{di})}
\textit{Keywords: directed, nonlinear, unsigned, bivariate, time-dependent.}

Directed information~\cite{massey1990causality,kramer1998directed} is a measure of information flow from a source time series $x$ to a target time series $y$ that is related to transfer entropy, but has no time lag between source and target \cite[App. C]{lizier2013book}.
The directed information can be computed as a difference between the conditional entropy of $y$ given its own past, and causally conditioned entropy~\cite{kramer1998directed,amblard_directed_2011,ziebart2013principle}.
For the same reasons as for causally conditioned entropy, we restrict its computation up to a history length of 10.
\begin{enumerate}[resume,noitemsep]
    \begin{multicols}{2}
        \item \code{di_gaussian}
        \item \code{di_kozachenko}
        \item \code{di_kernel_W-0.5}
    \end{multicols}
\end{enumerate}

\subsubsection{Stochastic interaction (\code{si})}
\textit{Keywords: undirected, nonlinear, unsigned, bivariate, time-dependent.}

Stochastic interaction~\cite{ay2015information,ay2003temporal} is a measure of integrated information between two processes in the context of their own past.
It is quantified by the difference between the joint entropy of the bivariate process and the individual entropies of each univariate process.
Both entropies are measured in context of (conditioned on) the history of the processes; we restrict this history to be only one-step, assuming a first-order Markov process.
\begin{enumerate}[resume,noitemsep]
    \begin{multicols}{2}
        \item \code{si_gaussian_k-1}
        \item \code{si_kozachenko_k-1}
        \item \code{si_kernel_W-0.5_k-1}
    \end{multicols}
\end{enumerate}

\subsubsection{Integrated information (\code{phi})}
\textit{Keywords: undirected, nonlinear, unsigned, bivariate, time-dependent.}

Integrated information was proposed to capture some aspects of consciousness as part of Integrated Information Theory (IIT)~\cite{tononi2004information,Tononi2016:IntegratedInformationTheory}.
We implement two proxy measures for IIT\,2.0~\cite{balduzzi2008integrated} from the \code{PhiToolbox}~\cite{PhiToolbox}:
\begin{itemize}
    \item $\Phi^*$ is a proxy of integrated information \cite{oizumi2016measuring}.
    It is an undirected measure that uses the concept of mismatched decoding in information theory~\cite{merhav1994information} and can be considered as the amount of loss of information due to the disconnection between two variables.
    \item $\Phi_G$ is a measure of integrated information derived from information geometry \cite{oizumi2016unified}.
    It is an undirected measure that quantifies the divergence between the actual probability distribution of a system and an approximated probability distribution where influences among elements are statistically disconnected.
    Here we implement the Gaussian-distribution model~\cite{oizumi2016measuring}.
\end{itemize}
Both measures are optionally normalized (divided) by entropy
(denoted by the modifier \code{norm}), yielding four SPIs:
\begin{enumerate}[resume,noitemsep]
    \begin{multicols}{2}
        \item \code{phi_star_t-1_norm-0}
        \item \code{phi_star_t-1_norm-1}
        \item \code{phi_Geo_t-1_norm-0}
        \item \code{phi_Geo_t-1_norm-1}
    \end{multicols}
\end{enumerate}

\subsection{Spectral (126 SPIs)}

Spectral SPIs are computed in the frequency or time-frequency domain, using either Fourier or wavelet transformations to derive spectral matrices.
Unless otherwise stated, the frequency-domain (i.e., Fourier-based) measures are computed using the \textit{Spectral Connectivity Toolbox}~\cite{eric_denovellis_2020_4088934}.
Each measure is based on the cross- and power-spectral densities at a given frequency and sampling rate, which is estimated via the multitaper technique.
All directed measures (excluding parametric spectral Granger causality) are nonparametrically estimated through spectral matrix factorization~\cite{dhamala_analyzing_2008,dhamala_estimating_2008}.
Moreover, the phase slope index is both computed in the frequency domain (via Fourier transformation) and the time-frequency domain (via Morlet transformation).
For more detail on the calculation, limitations, or interpretation of each of the following measures, see the review by Bastos and Schoffelen~\cite{bastos_tutorial_2016}.

Most methods in this literature category return a discrete set of values across a given frequency range.
Specifically, there is one value per frequency bin, $f$, for a given sampling frequency, $f_s$.
In order to compare processes with very different timescales in this work (e.g., economic time series sampled daily, or neural activity sampled at millisecond scale), we consider the time step, $\Delta t$ (s), between successive measurements of a time series to be rescaled by a timescale, $t_s$, appropriate for the process of interest, yielding a dimensionless time step, $\Delta \tilde{t} = \Delta t/t_s$.
Accordingly, we assume a sampling frequency $f_s = 1$ throughout (denoted by modifier \code{fs-1}).
We use 125 uniformly sampled bins across the entire frequency range, $f \in [f_0,f_s/2]$, where $f_s/2$ is the Nyquist frequency and $f_0 = 4/T$ is chosen as the minimal frequency for computational reasons.
In order to obtain SPIs, we take the mean (denoted by modifier \code{mean}) and the maximum (\code{max}) values of each measure over three ranges of the spectrum with corresponding modifiers, outlined below.
\begin{itemize}
    \item \code{fmin-0_fmax-0-5}: the full-frequency range, $f\in[0,f_s/2]$;
    \item \code{fmin-0_fmax-0-25}: the lower frequencies, $f\in[0,f_s/4]$; and
    \item \code{fmin-0-25_fmax-0-5}: the higher frequencies, $f\in[f_s/4,f_s/2]$.
\end{itemize}

\subsubsection{Coherence magnitude (\code{cohmag})} \label{sec:coherence}
\textit{Keywords: undirected, linear, unsigned, bivariate, frequency-dependent.}

The coherence (also known as coherence magnitude and ordinary coherence~\cite{schlogl_analyzing_2006} or coherence coefficient~\cite{bastos_tutorial_2016}) is an undirected frequency-dependent measure of linear association between time series $x$ and $y$.
Mathematically, it is the frequency domain equivalent of the squared time-domain cross correlation~\cite{bastos_tutorial_2016}.
We compute mean (denoted by \code{mean}) and maximum (\code{max}) summary statistics of the coherence for the three frequency ranges described above, giving six SPIs:
\begin{enumerate}[resume,noitemsep]
    \item \code{cohmag_multitaper_mean_fs-1_fmin-0_fmax-0-5} \label{spi:cohmag}
    \item \code{cohmag_multitaper_mean_fs-1_fmin-0_fmax-0-25}
    \item \code{cohmag_multitaper_mean_fs-1_fmin-0-25_fmax-0-5}
    \item \code{cohmag_multitaper_max_fs-1_fmin-0_fmax-0-5}
    \item \code{cohmag_multitaper_max_fs-1_fmin-0_fmax-0-25}
    \item \code{cohmag_multitaper_max_fs-1_fmin-0-25_fmax-0-5}
\end{enumerate}

\subsubsection{Coherence phase (\code{phase})} \label{sec:phase}
\textit{Keywords: undirected, linear, unsigned, bivariate, frequency-dependent.}

By omitting the magnitude operator of the coherence (Sec.~\ref{sec:coherence}), we obtain the complex-valued coherency, where the phase-difference angle (which we refer to as the coherence phase) has been used to infer a time-delayed dependence between two signals~\cite{bastos_tutorial_2016}.
We compute the coherence phase for the two summary statistics and three frequency ranges described above, yielding six SPIs:
\begin{enumerate}[resume,noitemsep]
    \item \code{phase_multitaper_mean_fs-1_fmin-0_fmax-0-5}
    \item \code{phase_multitaper_mean_fs-1_fmin-0_fmax-0-25}
    \item \code{phase_multitaper_mean_fs-1_fmin-0-25_fmax-0-5}
    \item \code{phase_multitaper_max_fs-1_fmin-0_fmax-0-5}
    \item \code{phase_multitaper_max_fs-1_fmin-0_fmax-0-25}
    \item \code{phase_multitaper_max_fs-1_fmin-0-25_fmax-0-5}
\end{enumerate}

\subsubsection{Group delay (\code{gd})}
\textit{Keywords: directed, linear, unsigned, bivariate, frequency-dependent.}

The group delay~\cite{gotman1983measurement} infers a directed, average time-delay between two signals by measuring the slope of the phase differences (coherence phase, Sec.~\ref{sec:phase}) as a function of the frequency (obtained through linear regression).
The slope is only computed for statistically significant coherence values and the time delay is obtained by a simple rescaling of the slope by $2\pi$.
We output the (rescaled) time-delay statistic for three frequency splits:
\begin{enumerate}[resume,noitemsep]
    \item \code{gd_multitaper_delay_fs-1_fmin-0_fmax-0-5}
    \item \code{gd_multitaper_delay_fs-1_fmin-0_fmax-0-25}
    \item \code{gd_multitaper_delay_fs-1_fmin-0-25_fmax-0-5}
\end{enumerate}

\subsubsection{Phase slope index (\code{psi})}
\textit{Keywords: directed, linear/nonlinear, unsigned, bivariate, frequency-dependent and time-frequency dependent.}

The phase slope index~(PSI)~\cite{nolte_robustly_2008} is a directed measure of information flow computed using the complex-valued coherency (see Sec.~\ref{sec:phase}).
Specifically, the phase slope index evaluates the consistency of the changes in phase differences across a pre-specified frequency range, weighted by the coherence.
Due to its availability in v0.23.0 of the \textit{MEG and EEG Analysis} python package~(MNE)~\cite{eric_larson_2021_4723951}, this is the only spectral measure that is computed in the time-frequency domain (denoted by modifier \code{wavelet}) in addition to the frequency domain (denoted by \code{multitaper}).
The time-frequency decomposition is given by a Morlet wavelet, with spectral densities indexed by a time-frequency tuple, given a sampling rate and number of cycles.
To obtain a statistic in the time-frequency domain, we take the average (denoted by the \code{mean} modifier) or the maximum (\code{max}) PSI across all time points.
We use the same pre-specified frequency ranges as outlined at the beginning of this section (i.e., low, high or full), giving three frequency-domain measures and six time-frequency domain measures due to the two summary statistics, i.e., nine SPIs in total:
\begin{enumerate}[resume,noitemsep]
    \item \code{psi_multitaper_mean_fs-1_fmin-0_fmax-0-5}
    \item \code{psi_multitaper_mean_fs-1_fmin-0_fmax-0-25}
    \item \code{psi_multitaper_mean_fs-1_fmin-0-25_fmax-0-5}
    \item \code{psi_wavelet_mean_fs-1_fmin-0_fmax-0-5_mean}
    \item \code{psi_wavelet_mean_fs-1_fmin-0_fmax-0-25_mean}
    \item \code{psi_wavelet_mean_fs-1_fmin-0-25_fmax-0-5_mean}
    \item \code{psi_wavelet_max_fs-1_fmin-0_fmax-0-5_max}
    \item \code{psi_wavelet_max_fs-1_fmin-0_fmax-0-25_max}
    \item \code{psi_wavelet_max_fs-1_fmin-0-25_fmax-0-5_max}
\end{enumerate}

\subsubsection{Imaginary coherence (\code{icoh})}
\textit{Keywords: undirected, linear, unsigned, bivariate, frequency-dependent.}

The imaginary part of the complex-valued coherency (referred to as imaginary coherence~\cite{bastos_tutorial_2016}) is argued as a means to obtain the coherence exclusively caused by a time delay (i.e., by removing instantaneous interactions that are present in the real axis)~\cite{nolte2004identifying}.
We compute the imaginary coherence for the three typical frequency ranges and two summary statistics, giving six SPIs:
\begin{enumerate}[resume,noitemsep]
    \item \code{icoh_multitaper_mean_fs-1_fmin-0_fmax-0-5}
    \item \code{icoh_multitaper_mean_fs-1_fmin-0_fmax-0-25}
    \item \code{icoh_multitaper_mean_fs-1_fmin-0-25_fmax-0-5}
    \item \code{icoh_multitaper_max_fs-1_fmin-0_fmax-0-5}
    \item \code{icoh_multitaper_max_fs-1_fmin-0_fmax-0-25}
    \item \code{icoh_multitaper_max_fs-1_fmin-0-25_fmax-0-5}
\end{enumerate}

\subsubsection{Phase locking value (\code{plv})}
\textit{Keywords: undirected, linear, unsigned, bivariate, frequency-dependent.}

The phase locking value (PLV)~\cite{lachaux1999measuring} is computed using the same formula as coherence, however applied to the amplitude-normalized Fourier transformed signals (i.e., normalized by individual tapers), which is argued to make it a more robust measure of phase synchronization than the coherence measures presented earlier~\cite{bastos_tutorial_2016}.
We compute the PLV for the three typical frequency ranges and two summary statistics, giving six SPIs:
\begin{enumerate}[resume,noitemsep]
    \item \code{plv_multitaper_mean_fs-1_fmin-0_fmax-0-5}
    \item \code{plv_multitaper_mean_fs-1_fmin-0_fmax-0-25}
    \item \code{plv_multitaper_mean_fs-1_fmin-0-25_fmax-0-5}
    \item \code{plv_multitaper_max_fs-1_fmin-0_fmax-0-5}
    \item \code{plv_multitaper_max_fs-1_fmin-0_fmax-0-25}
    \item \code{plv_multitaper_max_fs-1_fmin-0-25_fmax-0-5}
\end{enumerate}

\subsubsection{Pairwise phase consistency} \label{sec:ppc}
\textit{Keywords: undirected, linear, unsigned, bivariate, frequency-dependent.}

The pairwise phase consistency~(PPC)~\cite{vinck_pairwise_2010} measures phase synchronization by quantifying the distribution of the (per taper) phase differences, which means that, unlike PLV, it is not biased by the sample size~\cite{bastos_tutorial_2016}.
We compute the PPC for the three typical frequency ranges and two summary statistics, giving six SPIs:
\begin{enumerate}[resume,noitemsep]
    \item \code{ppc_multitaper_mean_fs-1_fmin-0_fmax-0-5}
    \item \code{ppc_multitaper_mean_fs-1_fmin-0_fmax-0-25}
    \item \code{ppc_multitaper_mean_fs-1_fmin-0-25_fmax-0-5}
    \item \code{ppc_multitaper_max_fs-1_fmin-0_fmax-0-5}
    \item \code{ppc_multitaper_max_fs-1_fmin-0_fmax-0-25}
    \item \code{ppc_multitaper_max_fs-1_fmin-0-25_fmax-0-5}
\end{enumerate}

\subsubsection{Phase lag index (\code{pli})}
\textit{Keywords: undirected, linear, unsigned, bivariate, frequency-dependent.}
\label{sec:pli}

The phase lag index (PLI)~\cite{stam2007phase} measures phase synchronization by averaging the sign of the (per taper) phase difference~\cite{bastos_tutorial_2016}.
In addition to computing the phase lag index (\code{pli}) for the frequency domain, we include weighted variations that make the measure more robust against electrophysical artifacts (namely, field spread), noise, and sample-size bias by weighting components based on the imaginary coherence~\cite{vinck2011improved}: the weighted phase lag index (\code{wpli}), the debiased squared phase lag index (\code{dspli}), and the debiased squared weighted phase lag index (\code{dswpli}).
In total, this gives 24 SPIs:
\begin{enumerate}[resume,noitemsep]
    \item \code{pli_multitaper_mean_fs-1_fmin-0_fmax-0-5}
    \item \code{pli_multitaper_mean_fs-1_fmin-0_fmax-0-25}
    \item \code{pli_multitaper_mean_fs-1_fmin-0-25_fmax-0-5}
    \item \code{pli_multitaper_max_fs-1_fmin-0_fmax-0-5}
    \item \code{pli_multitaper_max_fs-1_fmin-0_fmax-0-25}
    \item \code{pli_multitaper_max_fs-1_fmin-0-25_fmax-0-5}
    \item \code{wpli_multitaper_mean_fs-1_fmin-0_fmax-0-5}
    \item \code{wpli_multitaper_mean_fs-1_fmin-0_fmax-0-25}
    \item \code{wpli_multitaper_mean_fs-1_fmin-0-25_fmax-0-5}
    \item \code{wpli_multitaper_max_fs-1_fmin-0_fmax-0-5}
    \item \code{wpli_multitaper_max_fs-1_fmin-0_fmax-0-25}
    \item \code{wpli_multitaper_max_fs-1_fmin-0-25_fmax-0-5}
    \item \code{dspli_multitaper_mean_fs-1_fmin-0_fmax-0-5}
    \item \code{dspli_multitaper_mean_fs-1_fmin-0_fmax-0-25}
    \item \code{dspli_multitaper_mean_fs-1_fmin-0-25_fmax-0-5}
    \item \code{dspli_multitaper_max_fs-1_fmin-0_fmax-0-5}
    \item \code{dspli_multitaper_max_fs-1_fmin-0_fmax-0-25}
    \item \code{dspli_multitaper_max_fs-1_fmin-0-25_fmax-0-5}
    \item \code{dswpli_multitaper_mean_fs-1_fmin-0_fmax-0-5}
    \item \code{dswpli_multitaper_mean_fs-1_fmin-0_fmax-0-25}
    \item \code{dswpli_multitaper_mean_fs-1_fmin-0-25_fmax-0-5}
    \item \code{dswpli_multitaper_max_fs-1_fmin-0_fmax-0-5}
    \item \code{dswpli_multitaper_max_fs-1_fmin-0_fmax-0-25}
    \item \code{dswpli_multitaper_max_fs-1_fmin-0-25_fmax-0-5}
\end{enumerate}

\subsubsection{Directed transfer function (\code{dtf})}
\textit{Keywords: directed, linear, unsigned, bivariate, frequency-dependent.}
\label{sec:dtf}

The cross-spectral density matrix can be decomposed into a noise covariance matrix and a spectral transfer matrix, from which we obtain the directed transfer function~(DTF)~\cite{kaminski1991new}, which quantifies the inflow from $x \to y$ (according to the transfer matrix) normalized by the total inflow from all other signals into $y$ (the row-wise sum of the spectral transfer matrix).
In addition to the DTF, we include direct-DTF (\code{ddtf})~\cite{korzeniewska2003determination}, which extends DTF by conditioning out the influence of other signals (using the partialized cross-spectrum), giving 12 SPIs:
\begin{enumerate}[resume,noitemsep]
    \item \code{dtf_multitaper_mean_fs-1_fmin-0_fmax-0-5}
    \item \code{dtf_multitaper_mean_fs-1_fmin-0_fmax-0-25}
    \item \code{dtf_multitaper_mean_fs-1_fmin-0-25_fmax-0-5}
    \item \code{dtf_multitaper_max_fs-1_fmin-0_fmax-0-5}
    \item \code{dtf_multitaper_max_fs-1_fmin-0_fmax-0-25}
    \item \code{dtf_multitaper_max_fs-1_fmin-0-25_fmax-0-5}
    \item \code{ddtf_multitaper_mean_fs-1_fmin-0_fmax-0-5}
    \item \code{ddtf_multitaper_mean_fs-1_fmin-0_fmax-0-25}
    \item \code{ddtf_multitaper_mean_fs-1_fmin-0-25_fmax-0-5}
    \item \code{ddtf_multitaper_max_fs-1_fmin-0_fmax-0-5}
    \item \code{ddtf_multitaper_max_fs-1_fmin-0_fmax-0-25}
    \item \code{ddtf_multitaper_max_fs-1_fmin-0-25_fmax-0-5}
\end{enumerate}

\subsubsection{Directed coherence (\code{dcoh})}
\textit{Keywords: directed, linear, unsigned, bivariate, frequency-dependent.}

The directed coherence~\cite{baccala1998studying} is obtained from the inflow of $x\to y$ using the spectral transfer matrix (as per DTF, Sec.~\ref{sec:dtf}), normalized by their noise covariance.
We obtain six SPIs:
\begin{enumerate}[resume,noitemsep]
    \item \code{dcoh_multitaper_mean_fs-1_fmin-0_fmax-0-5}
    \item \code{dcoh_multitaper_mean_fs-1_fmin-0_fmax-0-25}
    \item \code{dcoh_multitaper_mean_fs-1_fmin-0-25_fmax-0-5}
    \item \code{dcoh_multitaper_max_fs-1_fmin-0_fmax-0-5}
    \item \code{dcoh_multitaper_max_fs-1_fmin-0_fmax-0-25}
    \item \code{dcoh_multitaper_max_fs-1_fmin-0-25_fmax-0-5}
\end{enumerate}

\subsubsection{Partial directed coherence (\code{pdcoh})}
\textit{Keywords: directed, linear, unsigned, bivariate, frequency-dependent.}

The partial directed coherence~\cite{baccala2001partial} from $x\to y$ is the inflow (as per DTF, Sec.~\ref{sec:dtf}) normalized by the total outflow from all other signals into $y$ (the column-wise sum of the spectral transfer matrix).
In addition to the partial directed coherence, we include the generalized partial directed coherence (\code{gpdoh})~\cite{baccala2007generalized}, which scales the relative strength of inflow from $x\to y$ by their noise covariance, yielding 12 SPIs:
\begin{enumerate}[resume,noitemsep]
    \item \code{pdcoh_multitaper_mean_fs-1_fmin-0_fmax-0-5}
    \item \code{pdcoh_multitaper_mean_fs-1_fmin-0_fmax-0-25}
    \item \code{pdcoh_multitaper_mean_fs-1_fmin-0-25_fmax-0-5}
    \item \code{pdcoh_multitaper_max_fs-1_fmin-0_fmax-0-5}
    \item \code{pdcoh_multitaper_max_fs-1_fmin-0_fmax-0-25}
    \item \code{pdcoh_multitaper_max_fs-1_fmin-0-25_fmax-0-5}
    \item \code{gpdcoh_multitaper_mean_fs-1_fmin-0_fmax-0-5}
    \item \code{gpdcoh_multitaper_mean_fs-1_fmin-0_fmax-0-25}
    \item \code{gpdcoh_multitaper_mean_fs-1_fmin-0-25_fmax-0-5}
    \item \code{gpdcoh_multitaper_max_fs-1_fmin-0_fmax-0-5}
    \item \code{gpdcoh_multitaper_max_fs-1_fmin-0_fmax-0-25}
    \item \code{gpdcoh_multitaper_max_fs-1_fmin-0-25_fmax-0-5}
\end{enumerate}

\subsubsection{Spectral Granger causality (\code{sgc})}
\textit{Keywords: directed, linear, unsigned, bivariate, frequency-dependent.}

Spectral Granger causality~\cite{granger1969investigating} is the frequency domain equivalent to Granger causality, computed via the spectral transfer matrix and noise covariance that are estimated using either a parametric (VAR model) approach~\cite{geweke_measurement_1982} or nonparametric (spectral factorization) approach~\cite{dhamala_analyzing_2008,dhamala_estimating_2008}.
We implement the nonparametric form from the \textit{Spectral Connectivity Toolbox}~\cite{eric_denovellis_2020_4088934} and the parametric implementation from v0.9 of \textit{NiTime}~\cite{rokem2009nitime}.
The VAR model is estimated via least squares using the same order for both processes, where the order is either inferred (denoted by modifier \code{order-None}) or fixed.
We infer a VAR model order from the Bayesian information criterion (with a maximum lag of 50, implemented in the \textit{NiTime} toolkit), as well as choosing fixed orders of one (denoted by \code{order-1}) and 20 (\code{order-20}).
The combination of parametric and nonparametric approaches, with each autoregressive order and summary statistic, gives 24 SPIs:
\begin{enumerate}[resume,noitemsep]
    \item \code{sgc_nonparametric_mean_fs-1_fmin-0_fmax-0-5}
    \item \code{sgc_nonparametric_mean_fs-1_fmin-0_fmax-0-25}
    \item \code{sgc_nonparametric_mean_fs-1_fmin-0-25_fmax-0-5}
    \item \code{sgc_nonparametric_max_fs-1_fmin-0_fmax-0-5}
    \item \code{sgc_nonparametric_max_fs-1_fmin-0_fmax-0-25}
    \item \code{sgc_nonparametric_max_fs-1_fmin-0-25_fmax-0-5}
    \item \code{sgc_parametric_mean_fs-1_fmin-0_fmax-0-5_order-None}
    \item \code{sgc_parametric_mean_fs-1_fmin-0_fmax-0-25_order-None}
    \item \code{sgc_parametric_mean_fs-1_fmin-0-25_fmax-0-5_order-None}
    \item \code{sgc_parametric_mean_fs-1_fmin-0_fmax-0-5_order-1}
    \item \code{sgc_parametric_mean_fs-1_fmin-0_fmax-0-25_order-1}
    \item \code{sgc_parametric_mean_fs-1_fmin-0-25_fmax-0-5_order-1}
    \item \code{sgc_parametric_mean_fs-1_fmin-0_fmax-0-5_order-20}
    \item \code{sgc_parametric_mean_fs-1_fmin-0_fmax-0-25_order-20}
    \item \code{sgc_parametric_mean_fs-1_fmin-0-25_fmax-0-5_order-20}
    \item \code{sgc_parametric_max_fs-1_fmin-0_fmax-0-5_order-None}
    \item \code{sgc_parametric_max_fs-1_fmin-0_fmax-0-25_order-None}
    \item \code{sgc_parametric_max_fs-1_fmin-0-25_fmax-0-5_order-None}
    \item \code{sgc_parametric_max_fs-1_fmin-0_fmax-0-5_order-1}
    \item \code{sgc_parametric_max_fs-1_fmin-0_fmax-0-25_order-1}
    \item \code{sgc_parametric_max_fs-1_fmin-0-25_fmax-0-5_order-1}
    \item \code{sgc_parametric_max_fs-1_fmin-0_fmax-0-5_order-20}
    \item \code{sgc_parametric_max_fs-1_fmin-0_fmax-0-25_order-20}
    \item \code{sgc_parametric_max_fs-1_fmin-0-25_fmax-0-5_order-20}
\end{enumerate}

\subsection{Miscellaneous (17 SPIs)}

A small number of methods do not fit squarely into any category listed above, and so we place them in a `miscellaneous' category.
Here, we discuss the use of linear and nonlinear model fits, for which we use \textit{scikit-learn}~\cite{pedregosa2011scikit}, cointegration, for which we use \textit{statsmodels}~\cite{seabold2010statsmodels}, and envelope correlation, for which we use MNE~\cite{eric_larson_2021_4723951}.

\subsubsection{Linear model fit (\code{lmfit})}
\textit{Keywords: directed, linear, unsigned, bivariate, contemporaneous.}

Linear regression is commonly used for establishing independence through model fits (e.g., see additive noise models in Sec.~\ref{sec:anm}).
As such, we use a number of linear models and record the mean squared error (MSE) of a regression of $y$ on $x$.
The following models (with the default parameters) from \textit{scikit-learn}~\cite{pedregosa2011scikit} are included: stochastic gradient descent regression with a squared loss (ordinary least squares fit) function (denoted by modifier \code{SGDRegressor}); Ridge regression, which uses $\ell_2$-norm regularization (\code{Ride}); the Elastic-Net model~\cite{friedman2010regularization,andersen2011interior}, which uses both $\ell_1$ and $\ell_2$-norm regularization (\code{ElasticNet}); and the Bayesian Ridge regressor uses a gamma distribution prior (with $\lambda_1 = \lambda_2 = 10^{-6}$) for the $\ell_2$-norm regularizor in Ridge regression (\code{BayesianRidge}).
This yields four SPIs:
\begin{enumerate}[resume,noitemsep]
    \begin{multicols}{2}
        \item \code{lmfit_SGDRegressor}
        \item \code{lmfit_Ridge}
        \item \code{lmfit_ElasticNet}
        \item \code{lmfit_BayesianRidge}
    \end{multicols}
\end{enumerate}

\subsubsection{Gaussian process model fit (\code{gpfit})}
\textit{Keywords: directed, nonlinear, unsigned, bivariate, contemporaneous.}

Similar to the linear model fits, we also use Gaussian process~\cite{williams1995gaussian} model fits as a nonparametric measure of influence of $x$ on $y$.
Here, we use a combination of kernels (with parameters chosen from default settings) from the \textit{scikit-learn} package~\cite{pedregosa2011scikit} and compute the MSE of their fit.
The fits are computed for the dot-product kernel with inhomogenity parameter $\sigma_0 = 1$ (\code{DotProduct}) and the radial basis function (RBF) kernel with length scale $l = 1$ (\code{RBF}).
Each of these kernels are separately combined with the constant kernel (with a constant of 1.0) and the white kernel (with a noise level of 1), giving two SPIs:
\begin{enumerate}[resume,noitemsep]
    \begin{multicols}{2}
        \item \code{gpfit_DotProduct}
        \item \code{gpfit_RBF}
    \end{multicols}
\end{enumerate}

\subsubsection{Cointegration (\code{coint})}
\textit{Keywords: undirected, linear, unsigned, bivariate, time-dependent.}

If two time series are individually integrated but some linear combination of them has a lower order of integration, then the series are said to be `cointegrated'~\cite{reinsel_elements_2003}.
We implement statistics for quantifying the cointegration of bivariate time series from two tests included in v0.12.0 of \textit{statsmodels}~\cite{seabold2010statsmodels}: the Augmented Engle-Granger (AEG, modifier \code{aeg})~\cite{engle1987co} two-step test and the Johansen test (\code{johansen})~\cite{pesaran1999autoregressive}.
\begin{itemize}
    \item For the AEG  test, we use a lag that is inferred via either AIC (modifier \code{aic}) or BIC (\code{bic}), with a maximum lag of 10 and obtain a $t$-statistic (\code{tstat}) of the unit-root test on the residuals.
    The time series is first detrended by assuming either a constant (\code{c}) or a constant and linear trend (\code{ct}).
    \item For the Johansen test, we output both the maximum eigenvalue (modifier \code{max_eig_stat}) and the trace (\code{trace_stat}) of the vector error correction model.
    Similar to the AEG test, we also assume a constant (\code{order-0}) or constant and linear trend (\code{order-1}) and fixed autoregressive lags of 1 (\code{ardiff-1}) and 10 (\code{ardiff-10}).
\end{itemize}
The combination of all tests and parameters gives 11 SPIs:
\begin{enumerate}[resume,noitemsep]
    \item \code{coint_johansen_max_eig_stat_order-0_ardiff-10}
    \item \code{coint_johansen_trace_stat_order-0_ardiff-10}
    \item \code{coint_johansen_max_eig_stat_order-0_ardiff-1}
    \item \code{coint_johansen_trace_stat_order-0_ardiff-1}
    \item \code{coint_johansen_max_eig_stat_order-1_ardiff-10}
    \item \code{coint_johansen_trace_stat_order-1_ardiff-10}
    \item \code{coint_johansen_max_eig_stat_order-1_ardiff-1}
    \item \code{coint_johansen_trace_stat_order-1_ardiff-1}
    \item \code{coint_aeg_tstat_trend-c_autolag-aic_maxlag-10}
    \item \code{coint_aeg_tstat_trend-ct_autolag-aic_maxlag-10}
    \item \code{coint_aeg_tstat_trend-ct_autolag-bic_maxlag-10}
\end{enumerate}

\subsubsection{Power envelope correlation (\code{pec})}
\textit{Keywords: undirected, linear, unsigned, bivariate, time-dependent.}

The envelope correlation~\cite{hipp2012large,khan2018maturation} is the correlation between the two amplitude envelopes of $x$ and $y$.
Power envelope correlation is computed using MNE, where we use six combinations of parameters, including whether the method was orthogonalized (modifier \code{orth}), the envelopes were squared and logs were taken prior to correlation (\code{log}), or the absolute of the correlation coefficient was used (\code{abs}):
\begin{enumerate}[resume,noitemsep]
    \begin{multicols}{2}
        \item \code{pec}
        \item \code{pec_orth}
        \item \code{pec_log}
        \item \code{pec_orth_log}
        \item \code{pec_orth_abs}
        \item \code{pec_orth_log_abs}
    \end{multicols}
\end{enumerate}

\section{Library of multivariate time series}

\label{sec:MTS-library}

This section summarizes the MTS assembled in our database.
The database (or `library' of MTS) includes synthetic time series, which we have generated, and real-world time series, which we have obtained from various publicly available resources.

MTS are described by $M$-vector processes, $\bm{Z} = (Z_1,\ldots,Z_M)$, where each $Z_i$ is a univariate stochastic process~\cite{reinsel_elements_2003}.
Here we assume a constant sampling period, $\Delta t$, such that each univariate time series in the tuple, $\bm{Z}$, can be represented by a discrete-time stochastic process, i.e., $Z_i = (Z_{i1},Z_{i2},\ldots)$.
The time series obtained from recording $T$ real-valued observations of these processes is referred to as a sample path (or realization) and is represented by lower-case variables, i.e., $\bm{z} = (z_1,\ldots,z_M)$, where each $z_i = (z_{i1},\ldots,z_{iT})$ and $z_{it} \in \mathbb{R}$.
Following conventional statistical notation, we use $\bm{Z}$ to denote a stochastic process, with $\bm{z}$ its realization and $\bm{z}_0 = (z_{i0},\ldots,z_{iM})$ the initial conditions.
The length of a realization, $\bm{z} = (\bm{z}_1,\ldots,\bm{z}_T)$, is denoted $T$ and the number of variables in a multivariate observation, $\bm{z}_1 = (z_{11},z_{21},\ldots,z_{M1})'$, is denoted $M$.

Time-series lengths, $T$, were chosen to be large enough to obtain reasonable statistics but small enough to allow the entire library to be processed within a feasible time limit: all datasets have a length of greater than $T = 100$ samples and less than $T = 2000$ samples.
The number of processes, $M$, was chosen to give enough data for correlating the SPIs (and, again, small enough to ensure we could process the library).
That is, for a given MTS, each SPI outputs $M(M-1)$ pairwise interactions that can be used to correlate this SPI with every other SPI (see details in Sec.~\ref{sec:similarity-stat}), and we need enough data to compute such a correlation.
The smallest multivariate system we consider has $M = 5$ processes (yielding $20$ pairwise interactions/samples for directed SPIs, and $10$ for undirected SPIs) and the largest is $M = 40$ (yielding $1560$ pairwise interactions/samples for directed SPIs, and $780$ for undirected SPIs).

Finally, certain high-dimensional dynamics that are recorded from real-world systems or generated from synthetic models can exhibit a low intrinsic dimensionality, meaning that the dynamics can be reproduced with a smaller number of variables than originally present in the signal.
For instance, fully synchronized oscillators appear as a set of processes that are all copies of one another, and so there is no more information in the entire system than the dynamics of one oscillator.
Datasets with low intrinsic dimensionality do not allow us to meaningfully capture similarities between different SPIs because the number of \textit{effective samples} in computing their similarity (as outlined above) is significantly lower than the original number of pairwise interactions (due to spatiotemporal correlation)~\cite{cliff2021assessing,bartlett_theoretical_1946}.
In analyzing the datasets for the purposes of inferring an empirical organization, we therefore discarded any MTS that requires less than three principal components to explain 95\% of the variance, ensuring that the \textit{effective sample size} is large enough to estimate the correlation, and that we only include systems that exhibit complex and non-trivial dynamics.

\subsection{Noise models}

We generated 96 datasets from uncorrelated noise (independently sampled from a distribution) and correlated noise (e.g., Brownian motion) models.
These models simulate dynamics for each process independently, and thus involve no interactions between processes.

\subsubsection{Uncorrelated noise}
\label{sec:uncorr-noise}

We generate a number of datasets using different distributions from the \textit{NumPy} package.
We perform this generation procedure for each of the following standard distributions:
\begin{itemize}
    \item \textit{Standard Normal distribution} (mean $\mu=0$ and standard deviation $\sigma=1$)
    \item \textit{Standard Cauchy distribution} (location $\mu=0$ and scale $\gamma=1$).
    \item \textit{Standard Exponential distribution} (rate parameter $\gamma=1$).
    \item \textit{Standard Gamma distribution} (shape $k=1$ and scale $\theta =1$).
    \item \textit{Standard $t$-distribution} (degrees of freedom $\mu=2$).
\end{itemize}
Each dataset was generated by sampling $T = [100,500,1000,2000]$ observations from $M = [5,10,20]$ independent distributions, giving 12 instances each (60 total).

\subsubsection{Correlated noise}

We used the Brain Dynamics Toolbox~\cite{heitmann2017handbook} to include stochastic differential equations for the following correlated noise models:
\begin{itemize}
    \item \textit{Ornstein--Uhlenbeck process}: generated using $M$ independent simulations of the differential equation: $dz_{it} = \theta (\mu-z_{it})\, dt + \sigma\,dW_{it}$.
    Parameters chosen were $\mu=0.5$, $\sigma=0.1$, $dt=T/10$, and with randomly generated initial conditions (with $z_{i0} \in [0,1]$) for each simulation.
    \item \textit{Arithmetic Brownian motion}: generated using $M$ independent simulations of the differential equation: $dz_{it} = \mu\, dt + \sigma\, dW_t$.
    Parameters chosen were $\mu=1$, $\sigma=1$, $dt=T/100$, and $\bm{z}_0=\bm{1}_M$ the vector of ones for each simulation.
    \item \textit{Geometric Brownian motion}: generated using $M$ independent simulations of the differential equation: $dz_{it} = \mu Z_{it}\, dt + \sigma z_{it}\, dW_{it}$.
    Parameters were chosen as $\mu = 0.5$, $\sigma = 0.5$, $dt = T/100$, and $\bm{z}_0 = \bm{1}_M$ for each simulation.
\end{itemize}
Each dataset is generated by sampling $T = [100,500,1000]$ observations from $M = [9,16,25]$ independent processes, giving 12 instances each (36 total).

\subsection{State-space models}

We generated 144 datasets from state space models, also known as vector autoregressive moving average (VARMA) models~\cite{reinsel_elements_2003}.
These dynamics are governed, in general, by an autoregressive component (up to order $p$) and a moving average component (up to order $q$):
\begin{equation}
    \bm{z}_{t} = \bm{a}_t + \sum_{i=1}^p \bm{\Phi}_i \bm{z}_{t-i} + \sum_{i=1}^q \bm{\Theta}_i \bm{a}_{t-i}\,,
\end{equation}
where the innovations $\bm{a}_t \sim \mathcal{N}(0,\bm{\Sigma})$ are normally distributed.
We randomly generate autoregressive matrices, $\bm{\Phi}_i$ and $\bm{\Theta}_i$, for simulating these dynamics.
Since it can be difficult to automate the generation of high-order VARMA models, we restrict our attention to first-order VAR/VMA.
For each type of system (VAR or VMA), we have eight configurations, given by varying each parameter below:
\begin{itemize}
    \item \textit{Coupling density}: either independent or randomly sampled coupling with a density of 10\%.
    \item \textit{Autocorrelation}: either low autocorrelation (coefficients of the matrices, $\bm{\Phi}$ and $\bm{\Theta}$, have magnitude in $[0,0.5]$) or high autocorrelation (coefficients have magnitude in $[0.5,1]$). The sign of the coefficients is randomly chosen to be positive or negative.
    \item \textit{Symmetry}: either symmetric or asymmetric coefficient matrices.
\end{itemize}
Each set of these configurations are generated with $T = [100,500,1000,2000]$ observations and $M = [5,10,20]$ processes.
Altogether, this gives 72 simulations from VAR models and 72 simulations from VMA models.

\subsection{Distributed dynamical systems}

In general, we consider three types of dynamical systems (with continuous states): coupled maps (discrete space and time), system of ordinary differential equations (discrete space, continuous time), and partial differential equations (continuous space and time).

\subsubsection{Coupled map lattice}

We generated 108 datasets from coupled map lattices, which are a popular model for studying spatiotemporal chaos.
Here, we use the model presented in~\cite{kaneko2011complex}, where each time series is generated from the following equation
\begin{equation}
    z_{it} = (1 -\epsilon) f(z_{it}) + \frac{\epsilon}{2}\left[f(z_{i+1,t}) + f(z_{i-1,t}) \right]\,,
\end{equation}
applied to the map, $f(z) = 1 - \alpha z^2$.
We generate a number of well-studied parameter choices with periodic boundary conditions (see~\cite{kaneko2011complex} for details), where the initial state is sampled from a uniform distribution, $z_{i0} \in [-0.5,1]$.
The qualitative classes of those parameters are listed below:
\begin{itemize}
    \item \textit{Frozen Chaos}: generated with $\alpha = 1.45$, $\epsilon = 0.2$ (with a transient period of 10,000 time points excluded).
    \item \textit{Pattern Selection}: generated with $\alpha = 1.71$, $\epsilon = 0.4$ (with a transient period of 10,000 time points excluded).
    \item \textit{Chaotic Brownian Motion of a Defect}: generated with $\alpha=1.85$, $\epsilon=0.1$ (with a transient period of 10,000 time points excluded).
    \item \textit{Defect Turbulence}: generated with $\alpha=1.895$, $\epsilon=0.1$ (with a transient period of 10,000 time points excluded).
    \item \textit{Spatiotemporal Intermittency (Type I)}: generated with $\alpha=1.7522$, $\epsilon=0.00115$, observed every 12 time steps (with no transient period).
    \item \textit{Spatiotemporal Intermittency (Type II)}: generated with $\alpha=1.75$, $\epsilon=0.3$ (with no transient period).
    \item \textit{Spatiotemporal Chaos}: generated with $\alpha=2.0$, $\epsilon=0.3$ (with a transient period of 1,000 time points excluded).
    \item \textit{Traveling wave}: generated with $\alpha=1.67$, $\epsilon=0.5$, observed every 2,000 time points (with a transient period of 10,000 time points excluded).
    \item \textit{Chaotic traveling wave}: generated with $\alpha=1.69$, $\epsilon=0.5$, observed every 5,000 time points (with a transient period of 10,000 time points excluded).
\end{itemize}
Each configuration was simulated with $M = [5,10,20]$ and $T = [100,500,1000,2000]$, and low-dimensional datasets were removed.

\subsubsection{Systems of ordinary differential equations (ODEs)}

Coupled oscillators are a type of coupled (nonlinear) ordinary differential equation (ODE) and popular distributed dynamical systems model exhibiting spaiotemporal phenomena, like synchrony and chaos.
Unless stated otherwise, the ODEs were simulated with $M=[5,10,20]$ processes and $T=[100,500,1000,2000]$ observations.
Note that some of these datasets were considered low-dimensional (according to our approximation of their intrinsic dimension) and so were not included in the final analysis.
\begin{itemize}
    \item The \textit{Kuramoto model}~\cite{kuramoto2003chemical} is generated using the \textit{pyclustering} library~\cite{Novikov2019}, which simulates the following equation:
    \begin{equation}
        \frac{d\theta_i}{dt} = \omega_{i} + \frac{K}{M} \sum_{j=1}^M a_{ij} \sin(\theta_j - \theta_i),
    \end{equation}
    where $\theta_i$ is the phase of oscillator $i$ at time $t$, $\omega_i$ is its natural frequency, $K$ is the coupling constant, and $a_{ij}$ is a boolean indicating whether oscillator $i$ is connected to oscillator $j$.
    In our library, we convert the phase, $\theta_i$, into a magnitude, i.e., $z_i = \sin{\theta_i}$, to avoid discontinuities.
    We use three different coupling schemes for $a_{ij}$: `all-to-all', where every oscillator is connected to every other oscillator; `bidirectional list', where each oscillator is connected with two neighbors in a list format; and `grid four', where each oscillator is connected to four of its neighbors in a grid format.
    For each coupling scheme, we use six different coupling strengths: $K = [-20,-4,-1,1,4,20]$, which affect the synchronization of the network (see~\cite{kuramoto2003chemical} for details).
    The combination of configurations, network size ($M$), and observation length ($T$), yielded 216 datasets, of which 130 were sufficiently high-dimensional to include in our analysis.

    \item The \textit{Kuramoto-Sakaguchi model}~\cite{kuramoto2003chemical} model extends the Kuramoto model to include phase shifts:
    \begin{equation}
        \frac{d\theta_i}{dt} = \omega_i + \frac{K}{M} \sum_{j=1}^M a_{ij} \sin(\theta_j - \theta_i - A_{ij})\,,
    \end{equation}
    where $A_{ij}$ is the phase frustration.
    Similar to the Kuramoto model, we use the filter $z_i = \sin{\theta_i}$, and three different coupling schemes: an all-to-all network; a circular network; and a random network.
    The phase frustration is set to the network adjacency (i.e., $A_{ij} = a_{ij}$).
    These models were simulated with $M=[9,16,25]$ processes and $T=[100,500,1000]$ observations via the \textit{Brain Dynamics Toolkit}~\cite{heitmann2017handbook}.

    \item The \textit{Stuart--Landau and Kuramoto model}~\cite{kuramoto2003chemical} combine the Stuart--Landau equations---representing a nonlinear oscillator near the Hopf bifurcation---with the Kuramoto model for synchronization.
    We use the \textit{pyclustering}~\cite{Novikov2019} implementation, which simulates the equation:
    \begin{equation}
        \frac{d\theta_i}{dt} = (i \omega_{i} + \rho^{2}_{i} - |\theta_i|^{2} )\theta_i + \frac{K}{M}\sum_{j=1}^{N} a_{ij}(\theta_j - \theta_i)\,,
    \end{equation}
    where $\rho_i$ is the radius, and $i \omega_i$ is the imaginary unit, $i$, multiplied by the natural frequency of that oscillator, $\omega_i$.
    We use the following (six) configurations:
    \begin{itemize}
        \item Two-way split radius (half the oscillators have $\rho_i=1$ and the other half have $\rho_i=2$), fixed natural frequency $\omega_i=1$, with coupling strength $K=1$, and an all-to-all network structure.
        \item Random radius $\rho_i \in [0,1]$, $K=1$, fixed natural frequency $\omega_i=1$, and an all-to-all network structure.
        \item Fixed radius $\rho_i=1$ for all oscillators, $K=1$, random natural frequency $\omega_i \in [0,1]$, all-to-all network structure.
        \item Fixed radius $\rho_i=1$, $K=1$, fixed $\omega_i=1$, with a bidirectional list structure.
        \item Fixed radius $\rho_i=1$, $K=1$, two-way split natural frequency (half with $\omega_i=1$, half with $\omega_i=2$), and an all-to-all network structure.
        \item Random radius $\rho_i\in[0,1]$, random natural frequency $\omega_i \in [0,1]$, $K=0.1$, and an all-to-all network structure.
    \end{itemize}

    \item The \textit{Hodgkin--Huxley oscillatory network}~\cite{chik2009selective} is a collection of interacting artificial neurons where the central element is based on the Hodgkin--Huxley model and the peripheral neurons take in some stimulus.
    We include four configurations that have been simulated using the \textit{pyclustering} toolbox~\cite{Novikov2019}, where we change only a small number of parameters that affect the synchronization of the peripheral neurons and central elements.
    We use the following (four) configurations:
    \begin{itemize}
        \item Two-way split stimulus with splits 25 and 47.
        We use the default parameters (see~\cite{Novikov2019}), except with \code{w1}, \code{w2}, and \code{w3} set to zero (intended to keep the oscillators unsynchonized).
        \item Two-way split stimulus with splits 25 and 27.
        We use the default parameters, except with \code{w1} set to 0.1, and \code{w2}/\code{w3} set to zero (intended to partially synchronize the oscillators).
        \item Two-way split stimulus with splits 25 and 47.
        We use the default parameters, except with \code{w1} set to 0.1, \code{w2} set to 5.0, and \code{w3} set to zero (intended to synchronize the oscillators).
        \item Three-way split stimulus with splits 25, 47, and 67.
        We use the default parameters, except with \code{deltah} set to 400.
    \end{itemize}
    \item A \textit{Morris--Lecar model} is an approximation to the original Hodgin--Huxley model for neuronal dynamics.
    We simulate a network of these neurons, with $M = [9,16,25]$ processes and $T = [100,500,1000]$ observations, via the \textit{Brain Dynamics Toolkit}~\cite{heitmann2017handbook, heitmann2018brain} (with default toolkit settings).

    \item The \textit{local-excitatory, global-inhibitory oscillatory network} (LEGION)~\cite{wang1995locally,wang1997image} is a class of synchronization models where each oscillator corresponds to a standard relaxation oscillator with two time scales.
    We use the \textit{pyclustering} implementation~\cite{Novikov2019}, where the relaxation oscillator is based on the van der Pol model.
    LEGION takes as input a number of parameters relating to biophysical processes, a stimulus for the neural network, and a network structure.
    We use the following six configurations:
    \begin{itemize}
        \item Stimulus of 1 for all oscillators, bidirectional list network structure, and otherwise default parameters.
        \item Three-way uniform-split stimulus (one-third with stimulus 0, next third with stimulus 1, remainder with stimulus 0), bidirectional list network structure, and otherwise default parameters.
        \item Three-way uniform-split stimulus, bidirectional list network structure, parameter \code{Wt} set to 4, parameter \code{fi} set to 10, and otherwise default parameters.
        \item Three-way random-split stimulus (same as uniform-split stimulus but the split locations are chosen randomly), bidirectional list network structure, parameter \code{Wt} set to 4, parameter \code{fi} set to 10, and otherwise default parameters.
        \item Three-way random-split stimulus, bidirectional list network structure, parameter \code{Wt} set to 4, parameter \code{fi} set to 0.8, and otherwise default parameters.
        \item Three-way random-split stimulus, bidirectional list network structure, parameter \code{teta_x} set to -1.1, and otherwise default parameters.
    \end{itemize}

    \item The \textit{hysteresis neural network}~\cite{zaghloul2012silicon}, implemented in the \textit{pyclustering} toolkit, comprises a network of neurons defined according to the following equations:
    \begin{align}
        \frac{dz_i}{dt} &= -z_i + \sum_{j=1}^M a_{ij} u_j\,, \\
        u_j &= h(z_j) =
        \begin{cases}
            +1 & \quad \textrm{for } z_j > -1\,, \\
            -1 & \quad \textrm{for } z_j < +1\,,
        \end{cases}
    \end{align}
    where $a_{ij}$ is the coupling between oscillators $i$ and $j$, $u_i$ are outputs, and $h(z_j)$ is a bipolar piecewise linear hysteresis.
    We generate dynamics using five configurations, each of which have an all-to-all network configuration and a fixed self-coupling $a_{ii} = -4$, while varying the neighbor coupling ($a_{ij}$ where $j \neq i$), the initial states $\bm{z}_0$, and the initial outputs $\bm{u}_0$:
    \begin{itemize}
        \item $a_{ij}=1$, a two-way split for the initial states (half of $\bm{z}_0$ set to 1, remainder set to 0), and all-positive outputs, $u_{i0} = 1$ for all $i$.
        \item $a_{ij}=-1$, a two-way split for the initial states (half of $\bm{z}_0$ set to 1, remainder set to 0), and all-positive outputs, $u_{i0} = 1$ for all $i$.
        \item $a_{ij}=-1$, random initial states, $\bm{z}_0 \in [0,1]$, all-negative outputs, $u_{i0} = -1$ for all $i$.
        \item $a_{ij}=1$, linearly increasing initial states, $z_{i0} = 1/(M-i)$, all-positive outputs, $u_{i0} = 1$ for all $i$.
        \item $a_{ij}=-1$, linearly increasing initial states, $z_{i0} = 1/(M-i)$, all-negative outputs, $u_{i0} = -1$ for all $i$.

    \end{itemize}
    \item A system of $M$ coupled \textit{van der Pol oscillators} is described by the equation:
    \begin{align}
         \frac{dz_i}{dt} &= u_i\,, \\
         \frac{du_i}{dt} &= \mu (1-z_i^2)u_i - z_i - \nu \sum_{j=1}^M a_{ij} \, z_i\,,
    \end{align}
    where $\mu$ and $\nu$ are scalar parameters, and $a_{ij}$ is the coupling strength between oscillator $i$ and $j$ (zero if there is no relationship).
    We use the \textit{Brain Dynamics Toolkit}~\cite{heitmann2017handbook,heitmann2018brain} to simulate the dynamics of the network of oscillators, with: \textit{(i)} a ring network structure, and \textit{(ii)} uniformly random $a_{ij} \in [0,1]$.

    \item The \textit{Wilson--Cowan} model describes the mean firing rates of reciprocally coupled excitatory and inhibitory neurons.
    We include simulations from a network of such oscillators from the \textit{Brain Dynamics Toolkit}~\cite{heitmann2017handbook,heitmann2018brain}, which solves the differential equations for excitatory, $u$, and inhibitory, $v$, neurons according to the following equation:
    \begin{align}
         \frac{du_i}{dt} &= -u_i + F\left(w_{uu} u_i - w_{uv} v_i + J_u + \sum_{j=1}^M a_{ij} u_i \right)\,, \label{eq:wc-1} \\
         \frac{dv_i}{dt} &= -v_i + F\left(w_{ie} u_i - w_{ii} v_i + J_i \right)\,,  \label{eq:wc-2}
    \end{align}
    where $a_{ij}$ is the coupling strength, $w_{uv}$ is the weight of the connection to $u$ from $v$, $J_u$ and $J_v$ are injection currents, and $F(v) = 1/(1+\exp(-v))$ is a sigmoid function (note we removed any variables from Eqs~\eqref{eq:wc-1} and \eqref{eq:wc-2} that were not used).
    For each of the simulations, we sample the mean firing rate of the excitatory neurons only: $z_i = u_i$.
    Moreover, we obtain the connectivity structure ($a_{ij}$) from the CoCoMac database (see~\cite{heitmann2017handbook}), and use the following parameters (obtained from code examples in~\cite{heitmann2017handbook}): $w_{uu} = 10$, $w_{vv} = -1$, $w_{uv} = 10$, $w_{vu} = 10$, $J_v = -2.5$, and $J_u = -8.5$.
\end{itemize}

\subsubsection{Partial differential equations}

Finally, partial differential equations represent spatiotemporally continuous systems that can be sampled discretely in space and time.
To obtain dynamics of this form, we simulated wave equations using the Brain Dynamics Toolkit~\cite{heitmann2017handbook,heitmann2018brain}.
The toolkit simulates these systems for both:
\begin{itemize}
    \item One spatial dimension using the equations,
    \begin{equation}
        \frac{\partial^2 z}{\partial t^2} = c^2 \, \frac{\partial^2 z}{\partial u^2}\,,
    \end{equation}
    where $c$ is a scalar value representing the wave propagation speed and $u$ is the spatial dimension; and
    \item Two spatial dimensions using the equations,
    \begin{equation}
        \frac{\partial^2 z}{\partial t^2} = c^2 \left( \frac{\partial^2 z}{\partial u^2} + \frac{\partial^2 z}{\partial v^2} \right)\,,
    \end{equation}
    where $u$ and $v$ are the two spatial dimensions.
\end{itemize}
For each system, we use $c=10$ (the default setting) and sample uniformly across space to obtain the MTS, $\bm{z}$.
Moreover, we use the default initial conditions, $\bm{z}_0$, which are proportional to a (one- or two-dimensional) Gaussian function with standard deviation based on the number of processes, $\sigma=M/20$.
The system is then solved using $M = [9,16,25]$ processes and $T = [100,500,1000]$ observations with periodic boundary conditions.

\subsection{Real-world data}

We obtained various real-world datasets from publicly available sources that cover geophysical, medical, physiological, financial, and other datatypes.
Many of these datasets were downloaded from the University of East Anglia (UEA) multivariate time-series classification database~\cite{ruiz2021great} (see this article for their references), where every task has a number of different classes, and often many instances of each class.
We attempt to minimize biasing towards certain tasks by using only ten instances of each class when including MTS from the UEA database.

\subsubsection{Medicine}

\begin{itemize}
    \item The UEA \code{ArticularyWordRecognition} datasets are electromagnetic articulographs of subjects speaking 25 words.
    Nine sensors were used in data collection with a sampling rate of 200\,Hz, giving $T = 144$ observations across $M = 9$ sensors.
    We include 10 instances from each of the 25 classes of this dataset, giving a total of 250 MTS for this datatype.

    \item The UEA \code{FaceDetection} datasets are 306-channel MEG recordings of subjects being shown either a face or a scrambled image.
    The data are recorded over 1.5\,s and then down-sampled to 250\,Hz and high-pass filtered at 1\,Hz, giving $T=62$ observations per 306 channels.
    We subsample the channels by randomly choosing a subsequence of channels, with $M \in [10,30]$.
    We include 10 instances from each of the two classes of this dataset, giving a total of 20 MTS for this datatype.

    \item The UEA \code{FingerMovements} datasets are 28-channel EEG recordings of subjects using a keyboard, intended for the classification task of predicting whether the subject was about to use their left or right hand.
    The data were downsampled to 100\,Hz, yielding $T = 50$ observations for each of the $M = 28$ channels.
    We include 10 instances from each of the two classes of this dataset, giving a total of 20 MTS for this datatype.

    \item The UEA \code{HandMovementDirection} datasets are 10-channel MEG recordings of subjects whilst moving a joystick towards one of four predetermined targets on a monitor.
    The trials have $T = 500$ observations of brain activity resampled at 400\,Hz across each of the $M = 10$ channels.
    We include 10 instances from each of the four classes of this dataset, giving a total of 40 MTS for this datatype.

    \item The UEA \code{Heartbeat} datasets are spectrograms of various locations (mainly the aortic, pulmonic, tricuspid, and mitral areas) of normal and abnormal heartbeats.
    The spectrogram is recorded over 5 s with a window size of 0.061\,s and an overlap of 70\%, where each process of the MTS is a frequency band from the spectrogram.
    Originally containing 61 processes representing different frequency bands, we subsampled the dataset to have between 15--25 processes.
    We include 10 instances from each of the two classes of this dataset, giving a total of 20 MTS for this datatype.

    \item The UEA \code{MotorImagery} datasets are recordings from an $8 \times 8$ ECoG platinum electrode grid, recorded at 1000\,Hz while subjects were imagining moving either the left small finger or the tongue.
    We subsampled the 64 electrodes by randomly choosing a subsequence of electrodes, with $M \in [10,30]$.
    We include 10 instances from each of the two classes of this dataset, giving a total of 20 MTS for this datatype.

    \item The \code{SelfRegulationSCP1} dataset records the 6-channel EEG signals of a subject, with the intention to classify if they are moving a cursor up and down a computer screen.
    The signals were recorded at a sampling rate of 256\,Hz for 3.5\,s, resulting in $T = 896$ observations per each of the $M = 6$ channels.
    We include 10 instances from each of the two classes of this dataset, giving a total of 20 MTS for this datatype.

    \item The UEA \code{SelfRegulationSCP2} dataset records the 7-channel EEG signals of an artificially respirated ALS patient, with the intention to classify if they are moving a cursor up and down a computer screen
    The signals were recorded at a sampling rate of 256\,Hz for 4.5\,s, resulting in $T = 1152$ observations per each of the $M = 7$ channels.
    We include 10 instances from each of the two classes of this dataset, giving a total of 20 MTS for this datatype.

    \item We obtained resting-state fMRI data of 100 subjects from the Human Connectome Project~\cite{van2012human}.
    The fMRI data consists of blood-oxygenated level-dependent (BOLD) signals from 333 parcels in the brain, which were minimally preprocessed and contain 1200 observations (see details in~\cite{shine2019human}).
    From this data, we randomly chose 10 subjects from which we used a subsequence of $M = 10$ parcels as our MTS.

    \item We obtained $N$-Back fMRI data of 100 subjects from the Human Connectome Project~\cite{van2012human}, which was minimally preprocessed and parcellated to give one 333-dimensional MTS with $T = 580$ observations per subject (see details in~\cite{shine2019human}).
    From this data, we randomly chose 10 subjects from which we used a subsequence of $M = 10$ parcels as our MTS.

    \item We obtained resting-state fMRI data from the UCLA Consortium for Neuropsychiatric Phenomics on both healthy individuals (130 subjects), and individuals with neuropsychiatric disorders, including schizophrenia (50 subjects), bipolar disorder (49 subjects), and attention deficit/hyperactivity disorder (ADHD, 43 subjects)~\cite{poldrack_phenome-wide_2016}.
    Each of the recordings had $T = 152$ observations and, from each of the four subject types (healthy, schizophrenia, biopolar disorder, and ADHD), we randomly selected 10 instances, and sampled $M = 15$ (consecutively numbered) brain regions from each instance.

    \item fMRI data was obtained for 100 anesthetized wild-type mice measured at rest~\cite{zerbi2021brain}.
    The data were parcellated using the Allen Common Coordinate Framework (CCF v3), and, similar to the human fMRI data, we randomly chose ten subjects, from which we subsampled MTS with $M = 10$ parcels and $T = 2000$ observations.

\end{itemize}

\subsubsection{Physiology}

\begin{itemize}
    \item The UEA \code{BasicMotions} dataset contains 3D accelerometer and 3D gyroscope recordings (at 10\,Hz for 10\,s) of students performing four activities
    We include 10 instances from each of the four classes of this dataset, giving a total of 40 MTS for this datatype.

    \item The UEA \code{Cricket} dataset~\cite{ko2005online} are 3D accelerometer readings (at 184\,Hz) of hand signals of cricket umpires performing twelve signals
    We include 10 instances from each of the twelve classes of this dataset, giving a total of 120 MTS for this datatype.

    \item The UEA \code{NATOPS} data contains 3D Euclidean coordinate data from sensors recorded at eight locations on the hands, elbows, wrists and thumbs, while the subject was performing six different gestures
    We include 10 instances from each of the six classes of this dataset, giving a total of 60 MTS for this datatype.

    \item The UEA \code{RacketSports} data contains 3D accelerometer and 3D gyroscope recordings (at 10\,Hz for 3\,s) of students playing a forehand/backhand in squash, or a clear/smash in badminton
    We include 10 instances from each of the four classes of this dataset, giving a total of 40 MTS for this datatype.
\end{itemize}

\subsubsection{Finance}

We obtained financial data on the daily open prices for various stocks and foreign exchange rates using the Pandas datareader~\cite{reback2020pandas}.
For each datatype, we downloaded time series for a random number of days, with $T\in[500,1500]$, between January 1 2010 to January 1 2020.

\begin{itemize}
    \item We downloaded five MTS of daily open prices from Yahoo Finance, each of which comprised $M\in[7,15]$ randomly sampled ticker symbols that made up the Dow--Jones Industrial Average index on August 31, 2020.

    \item We downloaded five MTS of daily open prices from Yahoo Finance, each of which comprised $M\in[7,15]$ randomly sampled ticker symbols that made up the SNP500 on August 31, 2020.

    \item We downloaded five MTS of daily open prices from the St.\ Louis FED (FRED), each of which comprised seven of the major foreign exchange rates: JPY/USD, USD/EUR, USD/CNY, USD/GBP, CAD/AUD, USD/AUD, CHF/USD, BRL/USD, and DKK/USD.
\end{itemize}

\subsubsection{Geophysics}

\begin{itemize}
    \item A number of hybrid (real-world and model-generated) MTS were downloaded from the CauseMe challenge~\cite{runge2020causality}.
    For these datasets, daily components were simulated using a random CMIP5 model, with random climate variables, that represent different climate subprocesses (thus, we refer to this dataset as a hybrid of real-world and simulated data).
    Each of the datasets listed below are named after the experiment names found on the website, with each originally comprising $M=40$ processes and $T=250$ observations.
    Here, we include 10 instances of each datatype, subsampling the processes to $M\in[10,20]$:
    \begin{itemize}
        \item \code{CLIM} is a climate-type dataset (featuring autocorrelation and time delays) that is aggregated to a monthly time resolution with dependencies up to a couple of months.
        \item \code{CLIMnoise} dataset is similar to \code{CLIM} but additionally contaminated with observational noise.
        \item \code{CLIMnonstat} dataset is similar to \code{CLIM} but additionally features nonstationarity.
        \item \code{WEATH} is a weather-type dataset (featuring autocorrelation, nonlinearity, and time delays) that has a 5-day time resolution with dependencies of up to a couple of months.
        \item \code{WEATHnoise} is similar to \code{WEATH} but additionally contaminated with observational noise.
        \item \code{WEATHsub} is similar to \code{WEATH} but is time-subsampled, i.e., only every third time step is observed.
    \end{itemize}

    \item Ten seismograms of recent earthquakes were downloaded from the IRIS Wilber 3 System~\cite{wilber3}.
    We downloaded seismic data from several (up to 25) stations that were closest to a given earthquake epicenter, with data recorded from the time of the earthquake until five minutes later.
    We randomly subsampled these datasets so that each of the ten MTS represented aligned seismic data from 15--25 stations, with between 1000--1500 observations in each dataset.

\end{itemize}

\subsubsection{Others}

\begin{itemize}
    \item The UEA \code{LSST} dataset is from the Photometric LSST Astronomical Time Series Classification Challenge to classify astronomical time-series data.
    The dataset contains simulations of the light curves of 14 astronomical objects across six different filters (passbands).
    We include 10 instances from each of the thirteen classes of this dataset, giving a total of 130 MTS for this datatype.

    \item The UEA \code{EigenWorms} dataset is intended to classify \textsl{C. Elegans} as either wild types or (one of four) mutant types based on time series recordings of their movements (which is translated via a projection onto a base shape known as eigenworms).
    We truncate the dataset to have $T=3000$ observations.
    We include 10 instances from each of the five classes of this dataset, giving a total of 50 MTS for this datatype.

    \item The UEA \code{PEMS-SF} data describes the occupancy rate (between 0 and 1) of different car lanes of San Francisco bay area freeways from Jan 1, 2008 to Mar 30, 2009.
    Each time series represents a single day sampled every 10 minutes (giving 144 observations) and the goal was to classify the correct day of the week.
    We subsample the channels by randomly choosing a subsequence of lanes (processes) with a random length $M \in [10,30]$.
    We include 10 instances from each of the seven classes of this dataset, giving a total of 70 MTS for this datatype.

    \item We obtained county-level COVID-19 cases in the USA, including records from 3006 counties over $T = 700$ days of the pandemic~\cite{nytimescovid}.
    From this database, we generated 10 instances of both the incidence (number of daily cases) and the cumulative incidence (cumulative sum of daily cases), where, in each instance, we randomly chose $M \in [7,15]$ counties.
\end{itemize}

\section{Supplementary Methods and Results}

\subsection{The empirical similarity index}
\label{sec:similarity-stat}

\begin{figure}[t!]
    \centering
     \includegraphics[width=.8\textwidth]{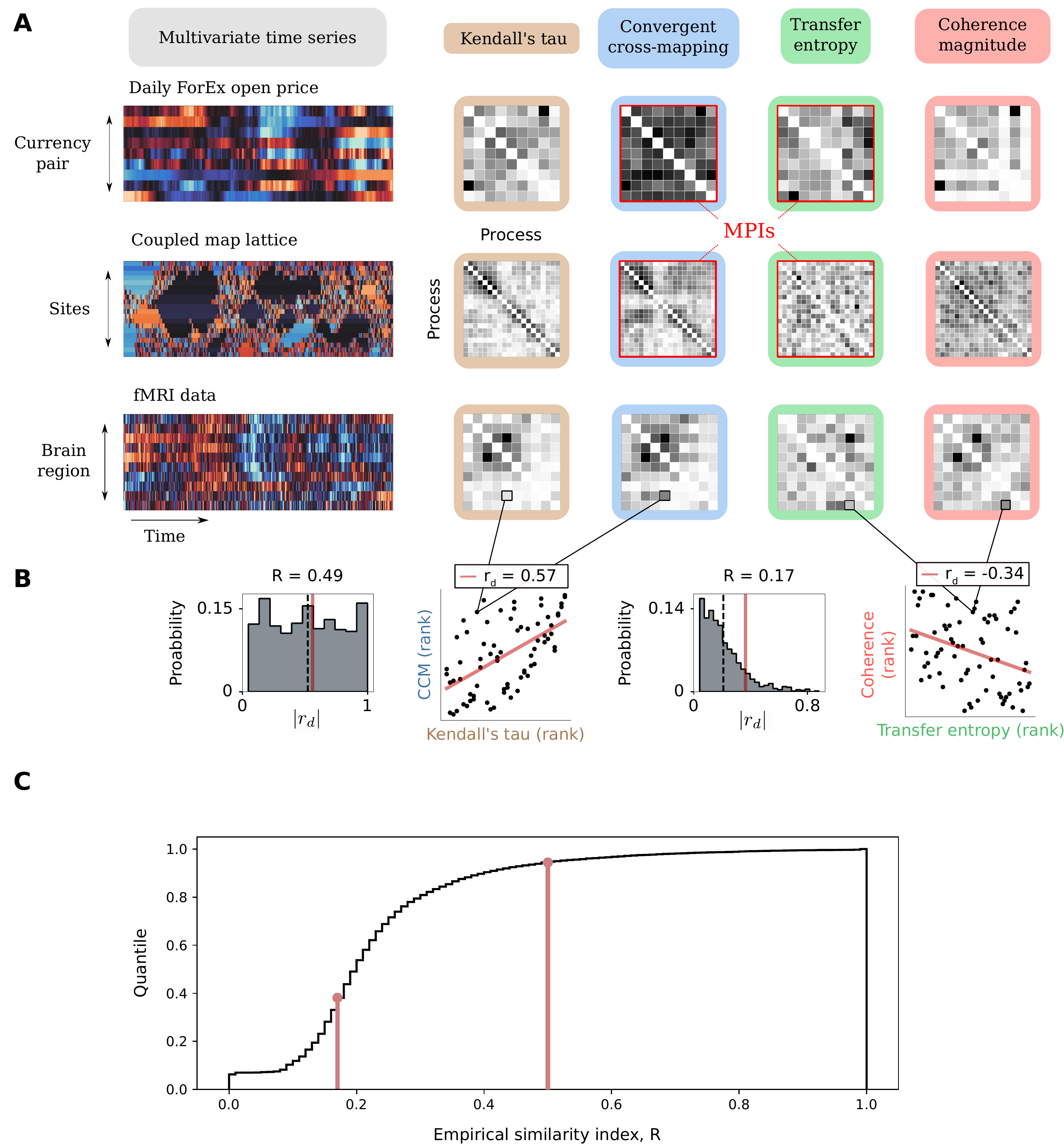}
    \caption{
    \textbf{We use the empirical similarity index, $R$, to measure the similarity between two SPIs, based on the Spearman's correlation coefficient between their pairwise matrices, averaged across all MTS.}
    \textbf{A.} From our MTS library, here we show three selected examples: \textit{(i)} exchange rates; \textit{(ii)} coupled maps; and \textit{(iii)} a functional Magnetic Resonance Imaging (fMRI) dataset.
    For each dataset, we also show how pairwise dependencies between all elements of the system can be captured as a matrix of pairwise interactions (MPI).
    The MPIs are normalized and shaded in grayscale, with white being the lowest value, and black being the highest.
    We show four selected examples from our library of SPIs:
    \textit{(i)} Kendall's $\tau$;
    \textit{(ii)} convergent cross-mapping;
    \textit{(iii)} transfer entropy; and
    \emph{(iv)} coherence magnitude.
    \textbf{B.} On a given dataset, we quantified the similarity between each pair of SPIs as the absolute Spearman's correlation coefficient of (non-diagonal) elements of the resulting pairwise matrices, as $|r_d|$, and averaged this quantity across all MTS to compute our empirical similarity score, $R$.
    Probability distributions of $|r_d|$ are shown for two SPI pairs: \textit{(i)} Kendall's $\tau$ and convergent cross-mapping, and \textit{(ii)} transfer entropy and coherence magnitude.
    The resulting $R$, the average of this distribution, is shown as a vertical dashed line.
    For each of these SPI pairs we also plot an example ranked scatter plot for the fMRI dataset (yielding an $|r_d|$ value annotated red in the distribution plot).
    \textbf{C.} The cumulative distribution function for the empirical similarity index across all pairs of SPIs, with annotations for: \textit{(i)} transfer entropy and coherence magnitude ($R = 0.17$); and \textit{(ii)} Kendall's $\tau$ and convergent cross-mapping ($R = 0.49$).
    }
    \label{fig:MPI_schematic}
\end{figure}

Some example MPIs are shown in Fig.~\ref{fig:MPI_schematic}A for SPIs based on transfer entropy (SPI \ref{spi:te}), coherence magnitude (SPI \ref{spi:cohmag}), Kendall's $\tau$ (SPI \ref{spi:kendalltau}), and convergent cross-mapping (SPI \ref{spi:ccm}) applied to three different MTS.
Different SPIs can yield quite different structures in their corresponding MPIs, and the level of similarity can vary across different datasets.
For example, in Fig.~\ref{fig:MPI_schematic}A, Kendall's $\tau$ and convergent cross-mapping exhibit markedly different behavior for the currency pair but qualitatively similar behavior on the coupled map lattice and human fMRI data.

The 1053 MTS have an average of approximately 11.89 processes, yielding 195\,112 pairwise interactions in total with which to compute the empirical similarity index per SPI pair.
Note, however, that undirected statistics have repeated entries in the upper and lower diagonals of the MPIs, and so the number of effective samples (in computing the correlation coefficient) is halved, which may induce a higher variance for undirected SPIs across all datasets when compared to directed SPIs~\cite{cliff2021assessing}.

The calculation of $r_d$ is depicted as ranked scatter plots in Fig.~\ref{fig:MPI_schematic}B for the similarity on human fMRI data between: \textit{(i)} Kendall's $\tau$~\cite{kendall1938new} and convergent cross-mapping (CCM) \cite{sugihara_detecting_2012}, $r_d = 0.57$; and \textit{(ii)} transfer entropy \cite{schreiber_measuring_2000} and coherence, $r_d = -0.34$.
For both of these pairs of SPIs, the distribution of $|r_d|$ across all 1053 MTS is also shown in Fig.~\ref{fig:MPI_schematic}B: the empirical similarity index between two SPIs, $R$, is then the average of this distribution, as $\langle |r_d| \rangle_d$.
While it was important to have a single summary of this similarity for a pair of SPIs, as the mean, it is important to note that some pairs of SPIs have quite a wide distribution of scores, $|r_d|$, across datasets, $d$, indicating that they can give highly correlated outputs on some MTS, but not on others.
An example of a wide distribution is shown in Fig.~\ref{fig:MPI_schematic}B, where the distribution of Kendall's $\tau$ and CCM is approximately uniformly distributed across MTS, ranging from some MTS for which the two SPIs are uncorrelated, $|r_d|\approx 0$, to other MTS for which there is perfect correlation, $|r_d| \approx 1$.

\begin{figure}[t!]
    \centering
    \includegraphics[width=\columnwidth]{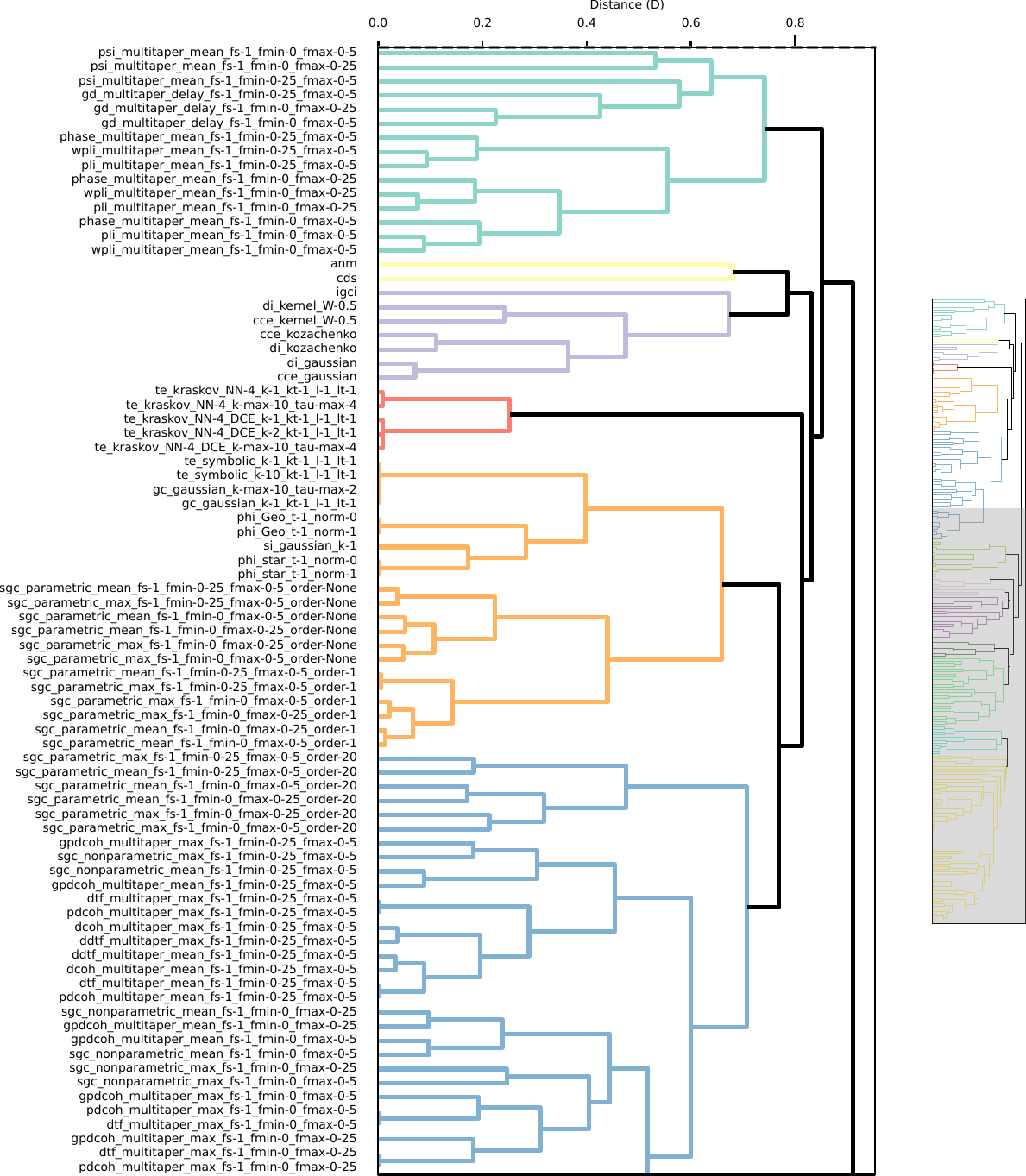}
    \caption{
    \textbf{A high-resolution dendrogram illustrates the empirical organization of SPIs at a fine-grained scale.}
    The color scheme illustrates clusters of SPIs below a cut-off of 0.76, and leaf nodes are labeled according to the SPI identifiers given in Sec~\ref{sec:spis-library}.
    }
    \label{fig:full_dendrogram}
\end{figure}

When using this approach to compare each pair of SPIs across the MTS library, we found a wide range of similarity indices, from $R \approx 0$ to $R \approx 1$, illustrated by the cumulative distribution function in Fig.~\ref{fig:MPI_schematic}C.
Each index was then used to construct the dendrogram (illustrated in Fig.~\ref{fig:full_dendrogram}) that gave us our findings in the main text.
While we were able to use our index to capture important empirical relationships using hierarchical clustering, it is difficult to interpret the raw index value, $R$, in terms of a statistical model (for, e.g., hypothesis testing).
This is because of two reasons: \textit{(i)} many MPIs contain values that are not completely independent of one another (due to higher-order dependencies), and \textit{(ii)} the set of all MPIs are themselves not completely independent (having occasionally overlapping dynamics).
As discussed in Sec.~\ref{sec:MTS-library}, the fact that the data used in computing the index are not independent may result in fewer effective samples than expected, which could bias the estimate of an absolute Spearman correlation and make many SPI pairs appear to have a higher correlation than they actually do.
Nevertheless, the effective sample size of the dataset does not affect empirical clustering.

\subsection{Modular versus literature categorization of SPIs}
\label{sec:module_vs_lit}

\begin{figure}[h]
    \centering
    \includegraphics[width=0.8\textwidth]{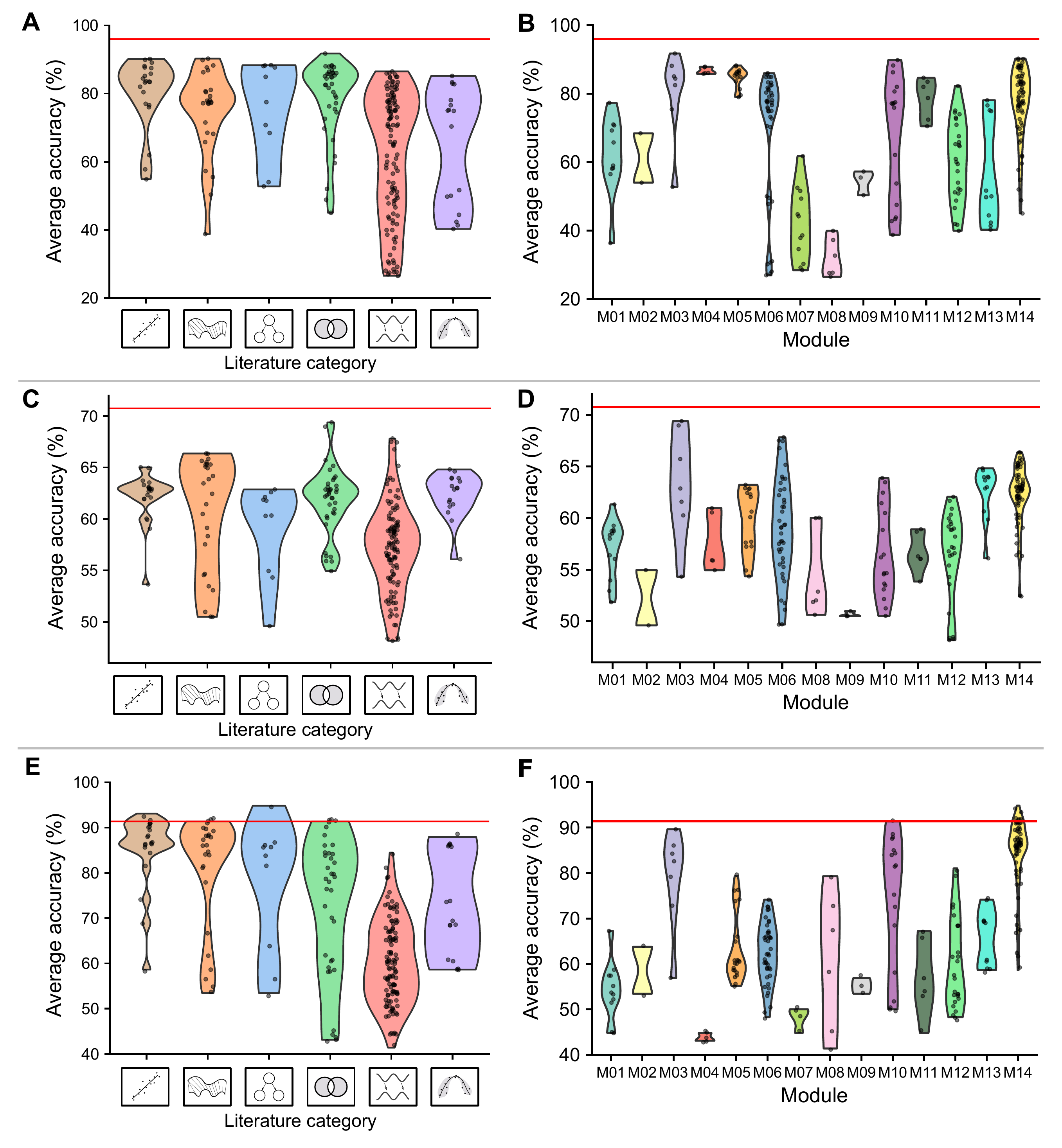}
    \caption{
    \textbf{Grouping SPIs by literature category can result in greater performance variability on a given MTS classification problem than grouping SPIs by modules.}
    Classification accuracy distributions over SPIs are shown as a violin plot when grouping SPIs as literature-derived categories, see Fig.~\ref{MT-fig:schematic} (left panels \textbf{A}, \textbf{C}, \textbf{E}) and by modular categories, see Fig.~\ref{MT-fig:dendrogram} (right panels, \textbf{B}, \textbf{D}, \textbf{F}).
    Results are shown for each datasets analyzed: smartwatch movement (\textbf{A}, \textbf{B}), EEG state (\textbf{C}, \textbf{D}), and fMRI film (\textbf{E}, \textbf{F}).
    Classification accuracy using the combination of all SPIs is shown as a horizontal red line.
    }
    \label{fig:literature_vs_module}
\end{figure}

We aimed to explore whether the literature categories or data-driven modules provided a more succinct representation of algorithm performance on the three MTS classification datasets studied in this paper.
Grouping the SPIs by the six literature categories (of Fig.~\ref{MT-fig:schematic}) revealed a wide variability of classification performance within each category.
By contrast, as shown in Fig.~\ref{MT-fig:classification}F, grouping SPIs by the fourteen data-driven modules (from Fig.~\ref{MT-fig:dendrogram}) results in far more characteristic performance levels within modules, suggesting our data-driven modular representation as a useful way to capture differential performance of SPIs on a given task.
For example, on the smartwatch activity dataset, SPIs from each literature category display a wide range of performance, including the information-theoretic SPIs (green in Fig.~\ref{fig:literature_vs_module}A), which vary in accuracy from 45\% (for the undirected SPI stochastic interaction with a Kozechenko--Leonenko estimator, \texttt{si\_kozachenko\_k-1}) to 92\% (for the directed SPI causally conditioned entropy with a Kozachenko--Leonenko estimator, \texttt{cce\_kozachenko}).
By contrast, we identify two modules containing consistently high-performing SPIs: M4 (transfer entropy) and M5 (parametric Granger causality and integrated information), in Fig.~\ref{fig:literature_vs_module}B.
These methods relate to measuring the communication and integration of information in a multidimensional process, adopting the Wiener--Granger theory of `causality' and `feedback' discussed earlier, suggesting that this way of conceptualizing and quantifying dependencies (predicated on self-predictability) is a useful one for analyzing these human behavioral recordings.
We find qualitatively similar results in the EEG state dataset (Figs~\ref{fig:literature_vs_module}C,D) and fMRI film dataset (Figs~\ref{fig:literature_vs_module}E,F).


\begin{figure}[t!]
    \centering
    \includegraphics[width=\columnwidth]{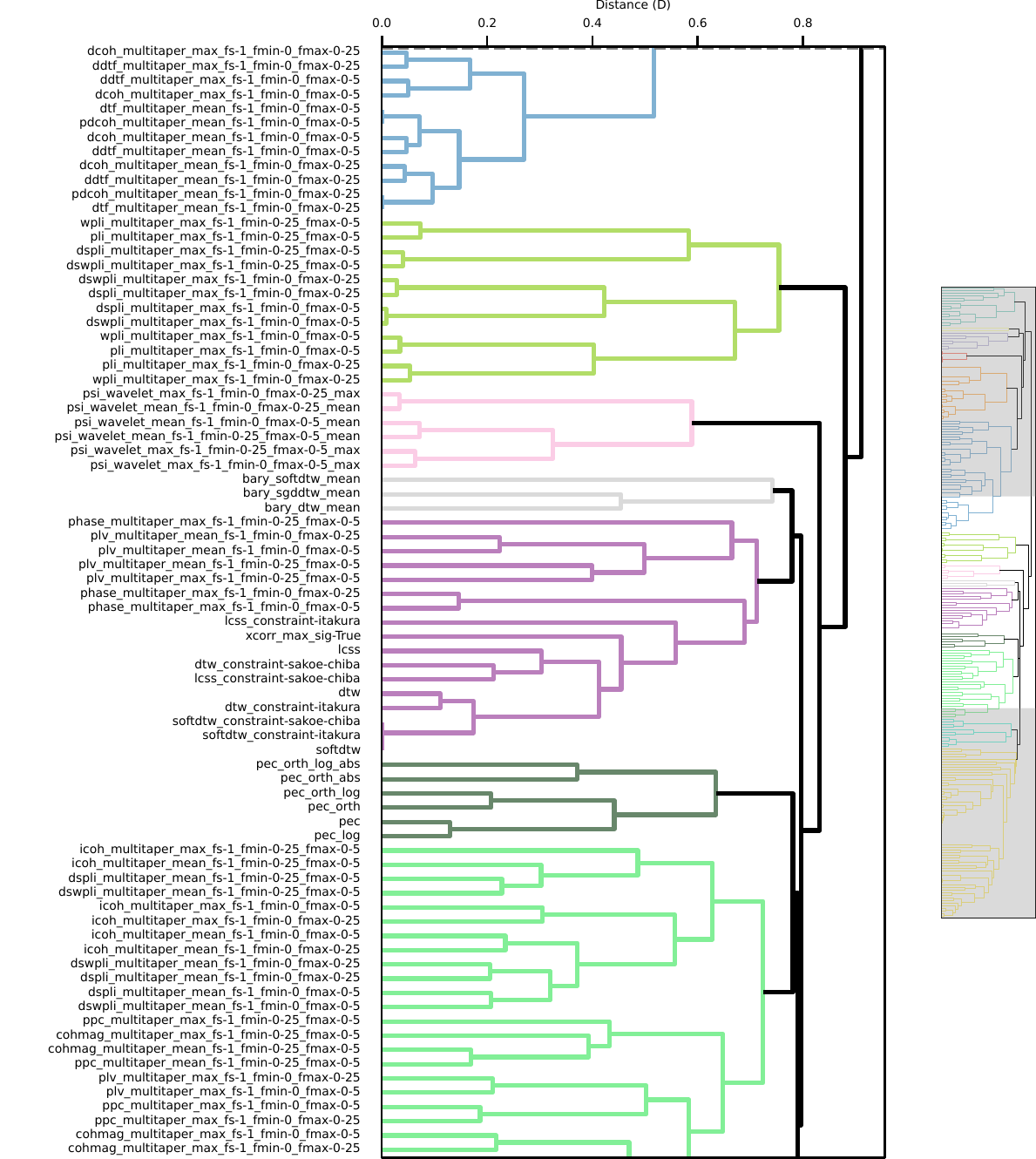}
\end{figure}

\begin{figure}[t!]
    \centering
    \includegraphics[width=\columnwidth]{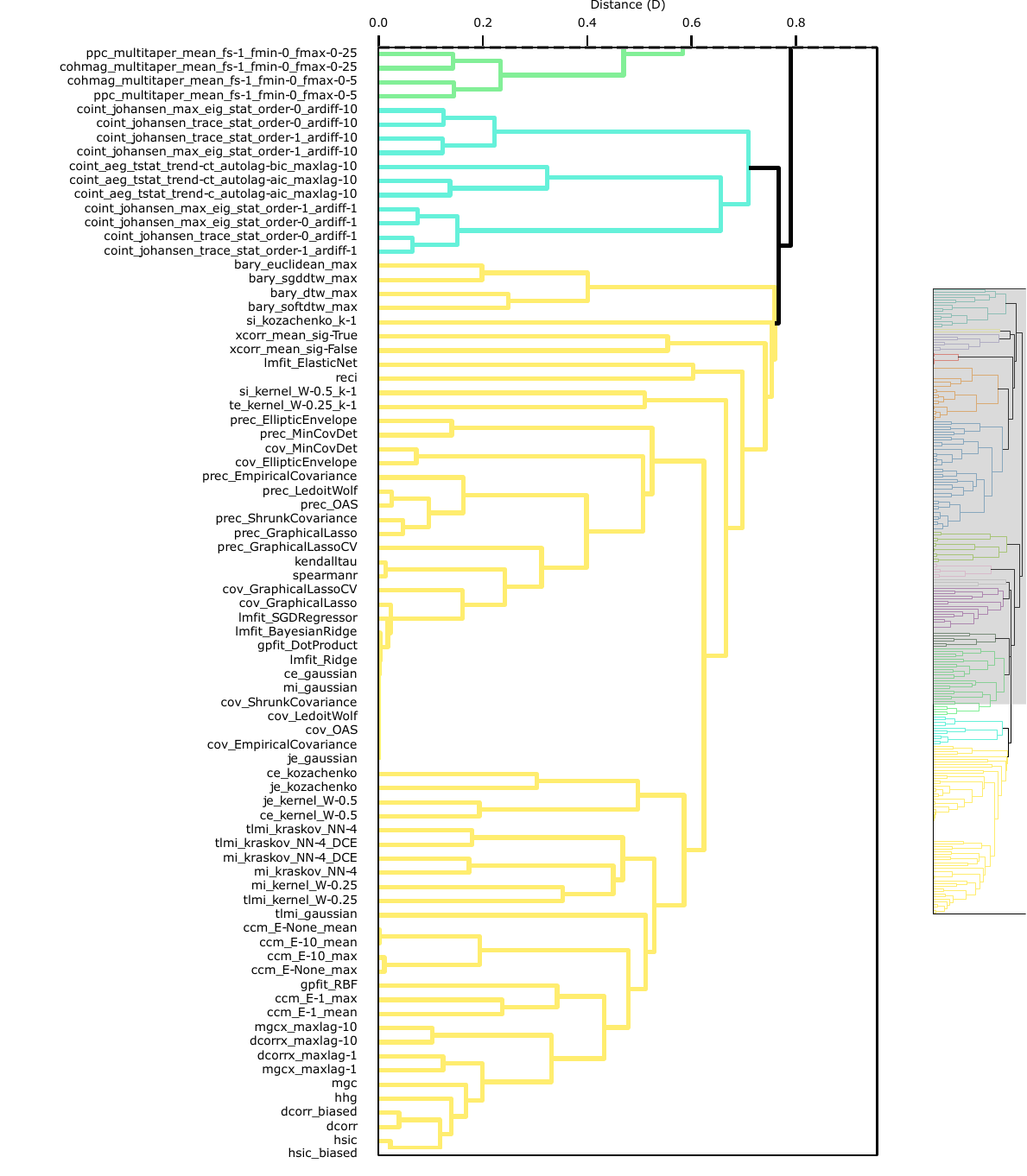}
\end{figure}

\clearpage

\bibliography{supplementary}
\bibliographystyle{benbibstyle}